\pdfoutput=1
\documentclass[acmsmall,screen,nonacm]{acmart}
\usepackage[utf8]{inputenc}
\usepackage[T1]{fontenc}
\usepackage[american]{babel}
\usepackage{mathtools}
\usepackage{thmtools}
\usepackage{xcolor}
\usepackage{paralist}
\usepackage{enumitem}
\usepackage[scale=0.85]{plex-mono}
\usepackage{stmaryrd}
\usepackage{tabularx}
\usepackage{mathpartir}
\usepackage{pftools}
\usepackage{listings} 
\usepackage[capitalise]{cleveref}
\usepackage{xspace}

\crefformat{section}{\S#2#1#3} 
\crefmultiformat{section}{\S#2#1#3}{ and~\S#2#1#3}{, \S#2#1#3}{, and~\S#2#1#3} 
\crefrangeformat{section}{\S#3#1#4 to~\S#5#2#6} 
\crefrangeformat{line}{Lines #3#1#4--#5#2#6}

\newcommand{\atool}[1]{\textsc{#1}}

\makeatletter
\def\operator#1{\@ifnextchar\bgroup {\operatorarg{\ensuremath{#1}}}{\ensuremath{#1}}}
\def\operatorarg#1#2{{#1}{\ensuremath{(#2)}}}
\def\spoperator#1#2{\@ifnextchar\bgroup{\spoperatorarg{\ensuremath{#1}}{\ensuremath{#2}}}{\ensuremath{#1}}}
\def\spoperatorarg#1#2#3{\ensuremath{#1#2#3}}
\makeatother

\newcommand{\NN}{\mathbb{N}}
\newcommand{\ZZ}{\mathbb{Z}}

\newcommand{\pto}{\rightharpoonup}
\newcommand{\tuple}[1]{\langle #1 \rangle}
\newcommand{\powersetof}[1]{\wp(#1)}
\newcommand{\set}[1]{\{#1\}}
\newcommand{\pset}[2]{\set{\,#1\mid#2\,}}
\newcommand{\dom}{\operatorname{\mathsf{dom}}}
\newcommand{\rng}{\operatorname{\mathsf{rng}}}

\newcommand{\lfp}{\operator{\mathsf{lfp}}}
\newcommand{\restrict}[2]{#1|_{#2}}
\newcommand{\ite}[3]{#1 \,?\, #2 \,:\, #3}

\newcommand{\var}{x}
\newcommand{\varp}{y}
\newcommand{\varpp}{z}
\newcommand{\fld}{f}
\newcommand{\flds}{\mathit{Fld}}
\newcommand{\vars}{\mathit{Var}}
\newcommand{\expr}{e}
\newcommand{\exprs}{\mathit{Exp}}
\newcommand{\com}{\mathit{com}}
\newcommand{\coms}{\mathit{Com}}
\newcommand{\stmt}{\mathit{st}}
\newcommand{\malloc}{\mathtt{malloc}}
\newcommand{\free}{\mathtt{free}}
\newcommand{\assume}{\mathtt{assume}}
\newcommand{\prog}{P}
\newcommand{\loc}{\ell}
\newcommand{\locp}{\loc'}
\newcommand{\initloc}{\loc_0}
\newcommand{\locs}{\mathit{Loc}}
\newcommand{\edges}{\Delta}
\newcommand{\edge}{\delta}

\newcommand{\addr}{a}
\newcommand{\addrp}{a'}
\newcommand{\addrs}{\mathit{Addr}}
\newcommand{\pnull}{\mathsf{null}}
\newcommand{\val}{v}
\newcommand{\vals}{\mathit{Val}}
\newcommand{\stack}{s}
\newcommand{\stackp}{\stack'}
\newcommand{\stacks}{\mathit{Stack}}
\newcommand{\fldvals}{\mathit{FldVal}}
\newcommand{\initfldvals}{\fldvals_0}
\newcommand{\heap}{h}
\newcommand{\heapp}{\heap'}
\newcommand{\heappp}{\heap''}
\newcommand{\pbheap}{\mathfrak{h}}
\newcommand{\stackheap}{\mathit{sh}}
\newcommand{\stackheapp}{\mathit{sh}'}
\newcommand{\heaps}{\mathit{Heap}}
\newcommand{\mfree}{\mathsf{free}}
\newcommand{\fldval}{\mathit{fm}}
\renewcommand{\state}{\sigma}
\newcommand{\statep}{\state'}
\newcommand{\stateset}{\Sigma}
\newcommand{\states}{\mathit{State}}
\newcommand{\fail}{\top}
\newcommand{\den}[1]{\llbracket #1 \rrbracket}
\newcommand{\eval}[2]{\den{#1}_{#2}}
\newcommand{\transformer}{F}

\newcommand{\step}{\rightarrow}

\newcommand{\post}{\mathsf{post}}
\newcommand{\postof}[1]{\post_{#1}}
\newcommand{\init}{\mathsf{init}}

\newcommand{\concdom}{\mathcal{C}}
\newcommand{\concord}{\sqsubseteq}
\newcommand{\concjoin}{\sqcup}
\newcommand{\bigconcjoin}{\bigsqcup}

\newcommand{\absdom}{\mathcal{A}}
\newcommand{\viewdesignation}{{\mathsf{v}}}
\newcommand{\viewstates}{\states^\viewdesignation}
\newcommand{\viewdom}{\absdom^\viewdesignation}
\newcommand{\vieword}{\sqsubseteq^\viewdesignation}

\newcommand{\bigviewmeet}{\bigsqcap}
\newcommand{\bigviewjoin}{\bigsqcup}

\newcommand{\view}{V}
\newcommand{\hinv}{H}

\newcommand{\viewgamma}{\gamma^\viewdesignation}
\newcommand{\viewalpha}{\alpha^\viewdesignation}
\newcommand{\viewpost}{\post^\viewdesignation}
\newcommand{\viewpostof}[1]{\viewpost_{#1}}
\newcommand{\viewtransformer}{\transformer^\viewdesignation}
\newcommand{\hinvcomp}{\triangleright}
\newcommand{\viewstep}{\step^\viewdesignation}
\newcommand{\push}{\mathsf{push}}

\newcommand{\mdom}{M}
\newcommand{\mop}{+}

\newcommand{\mBigOp}{\sum}
\newcommand{\mord}{\leq}
\newcommand{\mzero}{0}
\newcommand{\mval}{m}

\newcommand{\edgefn}{E}

\newcommand{\flowvar}{\operator{\mathit{fl}}}
\newcommand{\flow}{\operator{\mathit{flow}}}

\newcommand{\inflow}{\operator{\mathit{in}}}
\newcommand{\outflow}{\operator{\mathit{out}}}

\newcommand{\flowtransformer}{\operator{\mathit{Flow}}}

\newcommand{\flowdesignation}{\mathsf{f}}
\newcommand{\flowfld}{\mathsf{flow}}
\newcommand{\flowheap}{\heap^\flowdesignation}
\newcommand{\flowheapp}{\heap^{'\flowdesignation}}
\newcommand{\flowheaps}{\heaps^\flowdesignation}
\newcommand{\flowstates}{\states^\flowdesignation}
\newcommand{\flowdom}{\absdom^\flowdesignation}
\newcommand{\flowgamma}{\gamma^\flowdesignation}
\newcommand{\flowalpha}{\alpha^\flowdesignation}

\newcommand{\flowjoin}{\concjoin}

\newcommand{\bigflowjoin}{\bigconcjoin}

\newcommand{\flowviewdesignation}{{\flowdesignation\viewdesignation}}
\newcommand{\flowvieword}{\sqsubseteq^\flowviewdesignation}
\newcommand{\flowviewjoin}{\sqcup^\flowviewdesignation}

\newcommand{\bigflowviewjoin}{\bigsqcup}

\newcommand{\flowviewdom}{\absdom^\flowviewdesignation}
\newcommand{\flowviewalpha}{\alpha^\flowviewdesignation}
\newcommand{\flowviewgamma}{\gamma^\flowviewdesignation}
\newcommand{\flowviewtransformer}{\transformer^\flowviewdesignation}
\newcommand{\flowviewpost}{\post^\flowviewdesignation}
\newcommand{\flowviewpostof}[1]{\flowviewpost_{#1}}

\newcommand{\flowviewstep}{\step^\flowviewdesignation}
\newcommand{\sync}{\mathsf{sync}}
\newcommand{\syncd}{\mathsf{sync\_fp}}

\newcommand{\synccond}{\operator{\mathit{ViewLocal}}}
\newcommand{\synccondac}{\operator{\mathit{ViewLocalAcyc}}}
\newcommand{\gacyc}{\operator{\mathit{GlobalAcyclic}}}

\newcommand{\new}{\textcolor{blue}}

\lstdefinestyle{mycode}{
  language=c,
  morekeywords={assert,assume,define,Set,havoc},
  lineskip=.1em,
  numbers=left,
  xleftmargin=2em,
  numberstyle=\scriptsize,
  basicstyle=\small\ttfamily,
  keywordstyle=\color{teal}\ttfamily,
  commentstyle=\color{brown}\ttfamily,
  tabsize=2,
  firstnumber=last,
  numbersep=7pt,
  keepspaces=true,
  mathescape=true,
  moredelim=**[is][\color{purple}]{|<}{>|},
}
\usepackage{relsize}
\lstdefinestyle{inline}{
  keywords={},
}
\newcommand{\code}[2][]{\lstinline[style=inline,#1]!#2!}
\newcommand{\mymathtt}[1]{\text{\relscale{.9}\ttfamily#1}}
\newcommand{\mcode}[1]{\ensuremath{\mymathtt{#1}}}

\newcommand{\defineAuthor}[3]{
  \expandafter\newcommand\csname #1\endcsname[1]{%
    \ifdefined\finalversion{##1}%
    \else{\ifdefined\monochrome{\color{green!55!black}{##1}}%
      \else{\color{#3}##1}\fi}%
    \fi}
  \expandafter\newcommand\csname #1comment\endcsname[1]{%
    \ifdefined\finalversion{}%
    \else{\ifdefined\monochrome{}%
      \else{\color{#3}\csname #1\endcsname{#2: ##1}}\fi}%
    \fi}
  \expandafter\newcommand\csname #1out\endcsname[1]{%
    \ifdefined\finalversion{}%
    \else{\ifdefined\monochrome{\color{red!70!black}{\sout{##1}}}%
      \else{\color{#3}{\sout{##1}}}%
      \fi}%
    \fi}
  \expandafter\newcommand\csname #1mout\endcsname[1]{%
    \ifdefined\finalversion{}%
    \else{\ifdefined\monochrome{\color{red!70!black}{\text{\sout{\ensuremath{##1}}}}}%
      \else{\color{#3}{\text{\sout{\ensuremath{##1}}}}}%
      \fi}%
    \fi}
  \expandafter\newcommand\csname #1footnote\endcsname[1]{%
    \ifdefined\finalversion{}%
    \else{\ifdefined\monochrome{}%
      \else{\csname #1\endcsname{\footnote{\csname #1\endcsname{#2: ##1}}}}%
      \fi}%
    \fi}
}
\defineAuthor{tw}{TW}{purple}
\defineAuthor{sw}{SW}{red}
\defineAuthor{edg}{EG}{blue}
\defineAuthor{pr}{PR}{cyan}

\usepackage{tikz}
\usetikzlibrary{matrix,arrows,positioning,calc,fit,backgrounds,shapes}

\tikzset{%
  array/.style={matrix of nodes,nodes={draw, minimum size=5mm, anchor=center},column sep=-\pgflinewidth, row sep=-\pgflinewidth, nodes in empty cells,anchor=center},
  ptr/.style={*->, shorten <=-(1.8pt+1.4\pgflinewidth)},
  edge/.style={->, thick},
  dedge/.style={<->, dashed},
  fedge/.style={->, dashed},
  unode/.style={circle, draw=black, thick, minimum size=8mm, inner sep=0},
  mnode/.style={circle, draw=black, thick, fill=gray!20, minimum size=8mm, inner sep=0, font=\scriptsize},
  stackVar/.style={circle, fill=none, inner sep=0pt, minimum size=8mm, font=\normalsize, outer sep=-4pt},
  gnode/.style={circle, draw=black, thick, minimum size=8mm},
  pnode/.style={circle, draw=black, thick, minimum size=8mm},
  rnode/.style={draw=black, thick, minimum size=8mm},
  lbl/.style={circle, fill=none, inner sep=0pt, minimum size=8mm},
  dnode/.style={circle, draw=black, thick, dotted, minimum size=8mm},
  inflow/.style={circle, fill=none, inner sep=0pt, minimum size=5mm, font=\normalsize},
  phantomNode/.style={circle, fill=none, inner sep=0pt, minimum
    size=0pt}
}

\usepackage{pifont}
\newcommand{\symbolYes}{\smash{\ding{51}}\xspace}
\newcommand{\symbolNo}{\smash{\ding{55}}\xspace}

\newcommand{\aninv}{\mathsf{Inv}}

\begin{document}
\setdefaultenum{(i)}{(a)}{(a)}{(a)}

\title{Arithmetizing Shape Analysis}

\author{Sebastian Wolff}
\affiliation{%
  \institution{New York University}
  \country{USA}
}
\email{sebastian.wolff@cs.nyu.edu}

\author{Ekanshdeep Gupta}
\affiliation{%
  \institution{New York University}
  \country{USA}
}
\email{ekansh@nyu.edu}

\author{Zafer Esen}
\affiliation{%
  \institution{Uppsala University}
  \country{Sweden}
}
\email{zafer.esen@it.uu.se}

\author{Hossein Hojjat}
\affiliation{%
  \institution{TeIAS, Khatam University}
  \country{Iran}
}
\email{hojjat@ut.ac.ir}

\author{Philipp Rümmer}
\affiliation{%
  \institution{University of Regensburg}
  \country{Germany}
}
\affiliation{%
  \institution{Uppsala University}
  \country{Sweden}
}
\email{philipp.ruemmer@ur.de}

\author{Thomas Wies}
\affiliation{%
  \institution{New York University}
  \country{USA}
}
\email{wies@cs.nyu.edu}

\begin{abstract}
  Memory safety is an essential correctness property of software systems. For programs operating on linked heap-allocated data structures, the problem of proving memory safety boils down to analyzing the possible shapes of data structures, leading to the field of shape analysis. This paper presents a novel reduction-based approach to memory safety analysis that relies on two forms of abstraction: flow abstraction, representing global properties of the heap graph through local flow equations; and view abstraction, which enable verification tools to reason symbolically about an unbounded number of heap objects. In combination, the two abstractions make it possible to reduce memory-safety proofs to proofs about heap-less imperative programs that can be discharged using off-the-shelf software verification tools without built-in support for heap reasoning. Using an empirical evaluation on a broad range of programs, the paper shows that the reduction approach can effectively verify memory safety for sequential and concurrent programs operating on different kinds of linked data structures, including singly-linked, doubly-linked, and nested lists as well as trees.
\end{abstract}

\maketitle


\section{Introduction}

One of the most severe and common types of flaws in software systems are memory safety violations.
Memory safety violations typically happen in unsafe languages such as C/C++, for instance when the program tries to use a pointer to a memory location that has already been freed or that is out of bounds.
Memory safety has long been recognized as an important challenge in the development of secure software programs.
For instance, nearly 70\% of all Microsoft patches are fixes for memory safety bugs~\cite{Microbugs}, and
the US Cybersecurity and Infrastructure Security Agency (CISA) has recently urged software manufacturers to make it a top-level company goal to reduce and eventually eliminate memory safety vulnerabilities~\cite{cisa}.

In this paper, we focus on \emph{automatic} methods to prove the memory safety of programs operating on \emph{linked mutable data structures}. The key difficulty to verifying memory safety in this context is to determine the expected \emph{shape} of the data structures, a challenge that has led to the field of \emph{shape analysis}~\cite{DBLP:conf/popl/JonesM79,DBLP:journals/ftpl/ChangDMRR20} and a plethora of different methods; for instance, based on three-valued logic~\cite{DBLP:journals/toplas/SagivRW02}, automata techniques~\cite{DBLP:conf/cav/HolikLRSV13}, separation logic~\cite{DBLP:conf/tacas/DistefanoOY06, DBLP:conf/cav/BerdineCCDOWY07,DBLP:conf/sas/ChangRN07, DBLP:conf/cav/DudkaPV11}, bi-abduction~\cite{DBLP:conf/popl/CalcagnoDOY09,DBLP:conf/ecoop/HolikPRSVZ22}, and other tailor-made abstract domains~\cite{DBLP:conf/sas/PodelskiW05,DBLP:conf/popl/ChangR08,DBLP:conf/dagstuhl/AbdullaBCHJR08,DBLP:conf/cav/BouajjaniDERS10,DBLP:conf/sas/DudkaPV13,DBLP:journals/fmsd/IllousLR21,DBLP:conf/vmcai/ToubhansCR13,DBLP:conf/sas/GietRR23}. Today, the best tools competing in the \emph{MemSafety} category of the software verification competition SV-COMP are based on an intricate combination of shape analysis techniques (in particular abstract interpretation-based methods, symbolic execution, and model checking methods).

The mentioned approaches have in common that they generally need to be carefully tuned for a particular class of linked data structures to obtain good performance.
As a result, practical shape analyses often make trade-offs such as targeting only specific data structures (e.g., linked lists). Reasoning about data and concurrency is often treated as an afterthought or requires even more specialized abstract domains~\cite{DBLP:conf/cav/BouajjaniDERS10,DBLP:journals/pacmpl/0001W023,DBLP:conf/cav/Vafeiadis10}. The engineering effort involved in developing and maintaining these sophisticated analyses makes it difficult to integrate them with other techniques that target verification of properties unrelated to memory safety. In particular, modern software model checkers that aim to support the verification of general safety properties for imperative programs provide only very limited support for reasoning about pointers. It appears that they have been unable to benefit from the progress achieved in shape analyses techniques over the last two decades. State-of-the-art verification tools, even though they might perform very well on general verification tasks, can fail on even simple memory safety benchmarks involving singly-linked lists.

The goal of this paper is to provide an avenue for such tools to support shape reasoning without the need to integrate sophisticated shape analysis domains. To this end, we propose a reduction-based approach that can be implemented in a preprocessing step for static analyzers targeting heap-free programs. Concretely, we present an abstraction of heap-manipulating programs into integer programs that track shape information in auxiliary ghost variables. Assertions in the target program guarantee memory safety of the source program. In a sense, we effectively \emph{arithmetize} shape analysis for memory safety verification.

The reduction builds on recent advances on local reasoning techniques for heap-manipulating programs. Specifically, we combine ideas from the \emph{flow framework}~\cite{DBLP:journals/pacmpl/KrishnaSW18,DBLP:conf/esop/KrishnaSW20,DBLP:conf/tacas/MeyerWW23}, an approach based on separation logic for node-local reasoning about inductive properties of general heap graphs, as well as \emph{space invariants}~\cite{DBLP:conf/lpar/KahsaiKRS17}, which can summarize heap properties using node-local invariants. We formalize the soundness of our approach by deriving the target program as an abstract interpretation~\cite{DBLP:conf/popl/CousotC79,DBLP:conf/popl/CousotC77} of the source program. The resulting reduction is programmatic, data-structure-agnostic, and does not require the implementation of a sophisticated heap reasoning tool such as a prover for entailments in separation logic. 

As a result, we obtain a verification methodology that is extremely flexible:
\begin{inparaenum}
    \item by plugging in different flow domains, we can handle various classes of heap shapes and often many shapes with the same domain (e.g., lists, nested lists, and trees);
    \item since the final analysis is carried out by an off-the-shelve verification tool, our approach is agnostic of the considered data types or the control structure (e.g., recursion) present in programs; in particular
    \item the technique supports both sequential and concurrent heap-manipulating programs.
\end{inparaenum}

For evaluation, we have implemented our approach in a prototypical verification tool \atool{triceratops}, utilizing an off-the-shelf software model checker as back-end, and evaluate using a set of benchmarks taken from the SV-COMP~\cite{DBLP:conf/tacas/Beyer24}, as well as implementations of standard data structures written by ourselves. The benchmarks include both sequential and concurrent programs and cover a variety of different shapes, including singly-linked lists, doubly-linked lists, and trees. We find that \atool{triceratops} is able to verify memory safety effectively on such a diverse range of problems. In comparison with \atool{predatorHP}~\cite{DBLP:conf/tacas/PeringerSV20,DBLP:conf/hvc/HolikKPSTV16}, the 2024 SV-COMP gold medalist in the memory safety category \cite{DBLP:conf/tacas/Beyer24}, we observe that \atool{triceratops} tends to exhibit longer runtimes, but is able to cover a wider range of problems than the more specialized tool~\atool{predatorHP}.

The \emph{main contributions} of our paper are
\begin{inparaenum}
    \item the \emph{view abstraction} domain, an abstract domain inspired by the \emph{space invariants} approach~\cite{DBLP:conf/lpar/KahsaiKRS17} (\Cref{sec:view-abstraction});
    \item \emph{flow-based view abstraction,} a refinement of view abstraction with flow domains~\cite{DBLP:journals/pacmpl/KrishnaSW18,DBLP:conf/esop/KrishnaSW20,DBLP:conf/tacas/MeyerWW23} to capture heap shapes (\Cref{sec:flow-abstraction});
    \item an \emph{implementation} of our approach, resulting in the \atool{triceratops} tool (\Cref{sec:impl}); and
    \item an \emph{empirical evaluation} of the approach using a diverse set of sequential and concurrent programs (\Cref{sec:eval}).
\end{inparaenum}



\newcommand{\tprime}{\raisebox{1pt}{\ensuremath{'}}}

\section{Motivating Example}
\label{sec:motivation}

We consider the problem of proving \emph{memory safety} for heap-manipulating programs.
Specifically, we would like to show, fully automatically, the following properties:
\begin{enumerate}[leftmargin=*,labelindent=0pt,label={\bfseries (M\arabic*)}]
    \item
        Absence of unsafe accesses: read and write accesses to the heap only happen through valid pointers, i.e., through pointers that point to addresses that are allocated, have not yet been freed, and are not dangling (have received their addresses from the latest allocation).
        \label{cond:mem-safety-1}
    \item
        Absence of memory leaks: every portion of memory that has been allocated will eventually be freed.
        \label{cond:mem-safety-2}
    \item
        Absence of double frees: no portion of memory is freed twice.
        \label{cond:mem-safety-3}
\end{enumerate}
Memory safety is a fundamental correctness property of programs that is, however, often non-trivial to show, due to the need to understand the shape of heap-allocated data structures.

\begin{figure}
    \begin{minipage}{.3\textwidth}
    \begin{lstlisting}[gobble=6, mathescape=true]
      $\loc_0$: x := malloc;   $\label[line]{code:mot-ins:dummy-alloc}$
      $\loc_1$: while(*) {     $\label[line]{code:mot-ins:loop}$
            y := malloc;       $\label[line]{code:mot-ins:item-alloc}$
            z := x.l;          $\label[line]{code:mot-ins:next}$
            y.n := z;          $\label[line]{code:mot-ins:prepend}$
            x.l := y;          $\label[line]{code:mot-ins:move}$
        $\,$ }                 $\label[line]{code:mot-ins:end}$
        \end{lstlisting}
      \end{minipage}%
      \begin{minipage}{.33\textwidth}
      \begin{lstlisting}[gobble=6, mathescape=true]
      $\loc_2$: while(y != null) {   $\label[line]{code:mot-del:loop}$
            z := y.n;                $\label[line]{code:mot-del:next}$
            x.l := z;                $\label[line]{code:mot-del:move}$
            free(y);                 $\label[line]{code:mot-del:free}$
            y := z;                  $\label[line]{code:mot-del:trav}$
        $\,$ }                       $\label[line]{code:mot-del:end}$
      $\loc_3$: free(x);             $\label[line]{code:mot-del:free-dummy}$
      $\loc_4$:                      $\label[line]{code:mot-del:done}$
    \end{lstlisting}
  \end{minipage}%
  \begin{minipage}{.3\textwidth}
  \begin{tikzpicture}[>=stealth, font=\footnotesize, scale=0.8, every node/.style={scale=0.8}]
    \def\xsep{6}
    \def\ysep{3}

    \node[unode] (n1) {$\addr_1$};
    \node[unode, right=\xsep mm of n1] (n3) {$\addr_3$};
    \node[unode, right=\xsep mm of n3] (n2) {$\addr_2$};
    \node[mnode, right=\xsep mm of n2] (null) {$\pnull$};
    \node[mnode, above=\ysep mm of n1] (n4) {$\addr_4$};

    \node[stackVar, below=.4cm of n1] (x) {$\var$};
    \node[stackVar, below=.4cm of n3] (y) {$\varp,\varpp$};

    \draw[edge] (x) to (n1);
    \draw[edge] (y) to (n3);
    \draw[edge] (n1) to node[above] {$\mathtt{l}$} (n3);
    \draw[edge] (n4) to node[above] {$\mathtt{n}$} (n3);
    \draw[edge] (n3) to node[above] {$\mathtt{n}$} (n2);
    \draw[edge] (n2) to node[above] {$\mathtt{n}$} (null);
  \end{tikzpicture}    
  \end{minipage}
  \caption{A program $\prog$ that non-deterministically allocates a singly-linked list and then deallocates it again.
  The right side depicts a state reachable at $\loc_2$.\label{fig:example-program}}
  \end{figure}
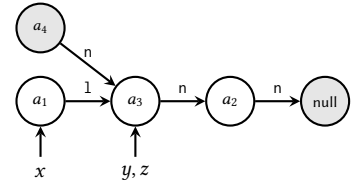

As the running example in this paper, we consider the case of a program allocating and later deallocating a singly-linked list, shown in \Cref{fig:example-program}. The program is written in a C-like language and uses functions \code{malloc} to allocate one memory cell and \code{free} to deallocate a previously allocated memory cell. We assume that the variables~\code{x}, \code{y}, \code{z} are initially set to \code{null}. The dummy object pointed to by \code{x}, allocated in \cref{code:mot-ins:dummy-alloc}, is responsible for storing a pointer~\code{x.l} to the first list node; initially, this pointer has value~\code{null} as well.
In the loop in \crefrange{code:mot-ins:loop}{code:mot-ins:end}, list nodes are allocated and added to the beginning of the list. For this, a new memory cell~\code{y} is allocated (\cref{code:mot-ins:item-alloc}), the next-pointer~\code{y.n} is set to the old value of \code{x.l} (\cref{code:mot-ins:prepend}), and the pointer~\code{x.l} is set to the address~\code{y} (\cref{code:mot-ins:move}).

In the loop in \crefrange{code:mot-del:loop}{code:mot-del:end}, the list nodes are deallocated again, starting with the head of the list. The loop first updates the pointer~\code{x.l} to the value~\code{x.l.n}, i.e., pointing to the second node in the list (\cref{code:mot-del:move}); it then deallocates the first list node (\cref{code:mot-del:free}), and finally sets the variable~\code{y} to the address of the new head (\cref{code:mot-del:trav}). Finally, in \cref{code:mot-del:free-dummy}, the dummy node~\code{x} is deallocated.

The graph on the right-hand side of \Cref{fig:example-program} illustrates a possible heap graph after one iteration of the second loop. The graph corresponds to a situation where object~$a_1$ has been allocated in \cref{code:mot-ins:dummy-alloc}, objects~$a_2, \ldots, a_4$ in \cref{code:mot-ins:item-alloc}, and the grey object~$a_4$ has been deallocated in the second loop.

Although simple, verifying the program automatically turns out to be surprisingly hard. To prove the memory safety properties~\ref{cond:mem-safety-1} to \ref{cond:mem-safety-3}, we need to observe that the program maintains the following two invariants.
\begin{enumerate}[leftmargin=*,labelindent=0pt,label={\bfseries (Inv\arabic*)}]
    \item The objects on the heap that have been allocated and not yet deallocated are exactly the objects reachable via the path~$\mcode{x.l.n}^*$ (or prefixes of this path).
    \item The pointers \code{x.l} and \code{y} coincide.
\end{enumerate}
Invariant~\textbf{Inv1}, in particular, is non-trivial to derive in an automatic way, as global reasoning about the shape of the heap graph is necessary.  This invariant can be derived using existing techniques for shape analysis, which require, however, carefully crafted shape domains to be efficient, and which are only precise for a certain set of shapes they are tuned for. Shape analysis techniques are often difficult to combine with other static analysis methods as well, for instance, to reason about both the shape and data (e.g., \cite{DBLP:conf/cav/BouajjaniDERS10,DBLP:conf/sas/GietRR23}). Because of that, many existing software model checkers and automatic verification tools will not be able to verify the program.

\paragraph{Contribution}
In this paper, we present a verification approach that carries out the required shape analysis in a \emph{completely local} way, and which is therefore easy to automate and easy to integrate with other automatic verification methods. Our approach is based on two main reasoning principles:
\begin{inparaenum}
    \item \emph{flow abstraction,} which reduces global properties of the heap to local flow equations that are required to hold for every object on the heap~\cite{DBLP:journals/pacmpl/KrishnaSW18,DBLP:conf/esop/KrishnaSW20,DBLP:conf/tacas/MeyerWW23};
    \item \emph{space invariants}~\cite{DBLP:conf/lpar/KahsaiKRS17}, here reformulated as \emph{view abstraction,} which model the contents of heap objects that are referenced by stack variables (\code{x}, \code{y}, \code{z} in the example) precisely, while representing all other objects on the heap using symbolic invariants.
\end{inparaenum}
The two approaches have in previous work been developed independently, and are in this paper combined to obtain a fully automatic approach for verifying memory safety of programs like the one in \Cref{fig:example-program}. It has to be noted that neither of the two approaches by itself is able to (automatically) verify the program in \Cref{fig:example-program}.

\paragraph{Flow abstraction}
To illustrate the reasoning carried out by our approach, imagine that we are associating natural numbers~$f_1, \ldots, f_4 \in \NN$ with the objects~$a_1, \ldots, a_4$, respectively, in \Cref{fig:example-program}. The numbers~$f_1, \ldots, f_4$, which we will later call the \emph{flow} of the heap graph, count the number of paths of the shape~\code{l.n}$^*$ leading to the respective object on the heap, and can be characterized by a set of simple equations, derived from the set of incoming \code{l}- or \code{n}-edges for each object:
\begin{align*}
    f_1 &= 0, & f_2 &= f_3, & f_3 &= 1 + f_4, & f_4 &= 0.
\end{align*}
The unique solution of the equations is~$f_1 = f_4 = 0$ and $f_2 = f_3 = 1$, and enables us to differentiate between nodes that are still reachable through the list ($a_2, a_3$, with flow~$> 0$) and objects that are outside of the list ($a_4$, with flow~$0$). As the entry point to the list, $a_1$ has itself flow $0$ but \emph{produces} flow $1$ for $a_3$. As a result, we can reformulate the invariant~\textbf{Inv1} as:
\begin{enumerate}[leftmargin=*,labelindent=0pt,label={\bfseries (Inv\arabic*\tprime)}]
    \item The objects on the heap that have been allocated and not yet deallocated are exactly the objects with flow~$>0$ (or currently pointed to by stack pointers).
\end{enumerate}

\paragraph{View abstraction}
We have not explained yet how flow can be computed in general, given a program heap with an unbounded number of objects. For this, we extend the data-structures on the heap by adding a field~\code{flow} to each object. In our example, each value~$f_i$ is represented by the field~$a_i\mcode{.flow}$. We can then \emph{instrument} the program to update the \code{flow} fields whenever heap objects or the heap graph change, and to ensure, using assertions, that objects are deallocated when their flow becomes $0$, and vice versa. To prove the absence of memory leaks, it is enough to assert that no objects with flow~$>0$ remain on program exit.

In order to infer invariants such as \textbf{Inv1\tprime},  we apply the concept of \emph{space invariants}~\cite{DBLP:conf/lpar/KahsaiKRS17} and aim at inferring a first-order predicate~$\aninv$ that holds for \emph{all} reachable states of \emph{all} objects ever allocated on the heap. The predicate~$\aninv$ can characterize the possible values of the \code{flow} field, and put those values in relationship with other fields of objects. To improve the precision of the encoding, we combine space invariants with a precise stack-based representation of the bounded number of objects that are currently pointed to by stack variables, resulting in the notion of \emph{view abstraction.}

We implement both flow and view abstraction using a source-to-source code transformation \emph{(instrumentation).} 
After instrumentation, we obtain a heap-less program that can be verified using off-the-shelf verification tools (e.g., software model checkers) by inferring two kinds of invariants: the space invariant~$\aninv$, as well as state invariants about the stack variables. Using an empirical evaluation on a broad range of programs, we show that our approach can effectively verify memory safety for programs operating on different kinds of linked data-structures, including lists and trees.

\subsection{Organization}

The rest of the paper is organized as follows. We first formally define syntax and semantics of the imperative language our analysis is targeting (\Cref{sec:prog-model}). After that, we introduce space invariants and view abstraction to symbolically reason about programs with an unbounded number of heap objects (\Cref{sec:view-abstraction}). We then define flow abstraction, we enables us to define global properties of the heap graph through local flow equations (\Cref{sec:flow-abstraction}). Our approach is implemented in the tool \atool{triceratops} (\Cref{sec:impl}), and evaluated on a set of benchmarks taken, among others, from the SV-COMP (\Cref{sec:eval}). We finally compare with related work (\Cref{sec:related-work}) and conclude the paper (\Cref{sec:conclusions}).



\section{Heap Programs}
\label{sec:prog-model}

In this section, we define an idealized imperative language of
heap-manipulating programs that we will use as the basis for
formalizing our static analysis.

We assume a finite set of \emph{program variables} $\vars = \vars_p \uplus \vars_d$ and a finite set of \emph{fields} $\fld \in \flds = \flds_p \uplus \flds_d$. The program variables are partitioned into pointer variables $\var_p \in \vars_p$ and data variables $\var_d \in \vars_d$, and similarly for fields.

Our formalization is parametric in a language of \emph{program expressions} $\expr \in \exprs = \exprs_p \uplus \exprs_d$, consisting of pointer expressions $\expr_p \in \exprs_p$ and data expressions $\expr_d \in \exprs_d$. We leave this expression language unspecified, but make some assumptions about its semantics that we specify below. The syntax of basic commands $\com \in \coms$ is then given by the following grammar:
\[
  \com \in \coms ::= \var_t := \varp_p.\fld_t \mid \var_p.\fld_t := \expr_t \mid \var_p := \malloc
  \mid \free(\var_p) \mid \var_t := \expr_t \mid \mathsf{assume}(\expr_d) 
\]
where $t \in \{p,d\}$. We omit type annotations whenever they are clear from context or irrelevant.
A \emph{program} $\prog$ is represented as a control flow automaton $\tuple{\locs,\edges,\initloc}$ that consists of a finite set of \emph{control locations} $\loc \in \locs$, a set of control flow edges $\edges \subseteq \locs \times \coms^* \times \locs$, and an initial control location $\initloc \in \locs$. We refer to a sequence of commands $\stmt \in \coms^*$ as a \emph{statement}.

\paragraph{Semantics.}
Let $\addr \in \addrs$ be a set of \emph{addresses}, which includes the dedicated address $\pnull$. The set of \emph{values} is $\vals = \addrs \uplus \ZZ$. A \emph{stack} $\stack \in \vars \to \vals$ provides a valuation for the program variables. A stack $\stack$ is \emph{safe} if it is well-typed, i.e., $\stack(\vars_p) \subseteq \addrs$ and $\stack(\vars_d) \subseteq \ZZ$. We denote by $\stacks$ the set of all safe stacks.
A \emph{heap} $\heap$ is a partial map that maps an address to a \emph{field valuation}, $\heap \in \addrs \pto \fldvals$ where $\fldvals = (\flds \uplus \set{\mfree}) \to \vals$. Note that a field valuation $\fldval \in \fldvals$ also assigns an auxiliary field $\mfree$ to a value $\fldval(\mfree) \in \ZZ$. Intuitively, $\malloc$ will never reuse a previously allocated address, even if that address has meanwhile been freed. Instead, freed addresses are simply marked as such using the free marker. We write $\fldval.\fld$ for $\fldval(\fld)$. We say that $\fldval$ is \emph{initial} if $\fldval(\flds_p) \subseteq \set{\pnull}$ and $\fldval(\mfree)=0$. We denote the set of all initial field valuations by $\initfldvals$.

This simpler, non-reallocating semantics coincides with the actual full, reallocating semantics, provided the program under the simpler semantics adheres to the following conditions \cite{DBLP:conf/vmcai/HazizaHMW16,DBLP:journals/pacmpl/MeyerW19,DBLP:journals/pacmpl/MeyerW20}:
\begin{inparaenum}
  \item dangling pointers are not dereferenced, and
  \item dangling pointers are not compared for equality with other pointers (except $\pnull$).
  \end{inparaenum}
The upshot of this design is that it simplifies checking the memory safety condition~\ref{cond:mem-safety-1} from \cref{sec:motivation}, which now becomes address-based:
our semantics utilizes the free markers to let programs that dereference dangling pointers fail.
The approach for pointer comparisons is similar, however, we refrain from making this precise in the formal development to avoid the need for explicitly defining the expression language.
For details on how our implementation checks both of these properties, refer to \cref{sec:impl:no-realloc}.

A heap $\heap$ is called \emph{safe} if it is well-typed and closed under following pointer fields, i.e. for all $\addr \in \dom(\heap)$, $\heap(\addr)(\flds_p) \subseteq \addrs \cap \dom(\heap)$ and $\heap(\addr)(\flds_d \uplus \set{\mfree}) \subseteq \ZZ$. We denote by $\heaps$ the set of all safe heaps. A pair $\stackheap = \tuple{\stack,\heap}$ is safe if $\stack$ and $\heap$ are safe, and $\rng(\stack) \cap \addrs \subseteq \dom(\heap)$. For the remainder of the paper, whenever we refer to stacks and heaps, we implicitly assume that they are safe, unless stated otherwise.
A program state $\state \in \states$ is of the form $\tuple{\ell,\stackheap}$ where $\loc$ is the control location of $\state$ and $\stackheap$
is a safe stack-heap pair. To define the semantics of basic commands, we assume that the expression language comes equipped with an evaluation function $\eval{\expr}{\stack}$ that evaluates an expression $\expr$ to a value under a given stack $\stack$. We require this function to be total and respect the types of expressions. That is, for all stacks $\stack$ and $\expr \in \exprs$, if $\expr \in \exprs_d$, then $\den{\expr}_\stack \in \ZZ$ and otherwise $\den{\expr}_\stack \in \stack(\vars_p) \cup \{\pnull\}$. The second condition ensures that the evaluation of pointer expressions does not generate new addresses out of thin air.

In the following, we assume that $\fail$ is an element distinct from all semantic objects. We use $\fail$ to indicate failure. For a set $S$ we write $S_\fail$ for $S \cup \set{\fail}$. The semantics of basic commands is given operationally in terms of a transition relation $\step \subseteq (\stacks \times \heaps)_\fail \times \coms \times (\stacks \times \heaps)_\fail$, written $\stackheap_\top,\com \step \stackheapp_\top$, which we define in \cref{fig:concrete-semantics}. It is easy to see that all stack and heap updates performed by $\step$ preserve safety. We extend $\step$ to a transition relation
$\step \subseteq (\stacks \times \heaps) \times \coms^* \times (\stacks
\times \heaps)_\fail$ on statements in the expected way.

The concrete domain $\concdom$ is the \emph{topped} powerset domain of states, $\concdom = \powersetof{\states}_\fail$, with the partial order $\concord$ defined by $\stateset_\fail \concord \stateset'_\fail$ iff $\stateset_\fail,\stateset'_\fail \in \powersetof{\states}$ and $\stateset_\fail \subseteq \stateset'_\fail$ or $\stateset'_\fail = \fail$. We lift the
transition relation to a strongest postcondition transformer
$\post \in \edges \to \concdom \to \concdom$:
\[\postof{\tuple{\loc,\stmt,\locp}}(\stateset_\fail) = \pset{\tuple{\locp,\stackheapp}}{\exists \tuple{\loc,\stackheap} \in \stateset_\fail.\, \stackheap,\stmt \step \stackheapp} \concjoin \bigconcjoin \pset{\fail}{\exists \tuple{\loc,\stackheap} \in \stateset_\fail.\, \stackheap,\stmt \step \fail}\enspace.\]
A state $\state$ is called \emph{initial} if $\state=\tuple{\initloc,\tuple{\stack_0,\heap_0}}$ where $\stack_0(\vars_p) \subseteq \set{\pnull}$ and $\heap_0 = \{ \pnull \mapsto \fldval[\mfree \mapsto 1]\}$ for some $\fldval \in \initfldvals$. The set of all initial states is denoted by $\init$.
The semantics $\den{\prog} \in \concdom$ of a program $\prog$ is then defined as the least fixpoint of $\post$ over $\edges$, starting from $\init$:
\[
\den{\prog} = \lfp{\transformer} \quad \text{where} \quad \transformer = \lambda \stateset.\, \bigconcjoin_{\edge \in \edges} \postof{\edge}(\stateset) \concjoin \init\enspace.
\]
We say that $\prog$ is \emph{memory safe} if $\den{\prog} \neq \fail$.

\begin{figure}[t]
\def \MathparLineskip {\lineskip=1.3ex}
\begin{mathpar}
  \inferH{Assume}
  {\eval{\expr}{\stack} \neq 0}
  {{\tuple{\stack,\heap}},{\assume(\expr)} \step {\tuple{\stack,\heap}}}
  \and
  \axiomHtop{Assign}
  {{\tuple{\stack,\heap}},{\var := \expr} \step {\tuple{\stack[\var \mapsto \eval{\expr}{\stack}],\heap}}}
  \\
  \inferH{Read}
  {\stack(\varp) = \addr\\
    \heap(\addr).\mfree = 0
  }
  {{\tuple{\stack,\heap}},{\var := \varp.\fld} \step \tuple{\stack[\var \mapsto \heap(\addr).\fld],\heap}}
  \and
  \inferH{ReadAbort}
  {\stack(\varp) = \addr\\
    \heap(\addr).\mfree \neq 0
  }
  {{\tuple{\stack,\heap}},{\var := \varp.\fld} \step \fail}
  \and
  \inferH{Write}
  {\stack(\var) = \addr\\
    \heap(\addr).\mfree = 0\\
    \eval{\expr}{\stack} = \val
  }
  {{\tuple{\stack,\heap}},{\var.\fld := \expr} \step \tuple{\stack,\heap[\addr \mapsto \heap(\addr)[\fld \mapsto \val]]}}
  \and
  \inferH{WriteAbort}
  {\stack(\var) = \addr\\
    \heap(\addr).\mfree \neq 0
  }
  {{\tuple{\stack,\heap}},{\var.\fld := \expr} \step \fail}
  \\
  \inferH{Malloc}
  {\addr \notin \dom(\heap)\\
    \fldval \in \initfldvals
  }
  {{\tuple{\stack,\heap}},{\var := \malloc} \step \tuple{\stack[\var \mapsto \addr],\heap[\addr \mapsto \fldval]}}
  \qquad
  \and
  \axiomHtop{Abort}
  {\fail,\com \step \fail}
  \qquad\qquad
  \and
  \inferH{Free}
  { \stack(\var) = \addr\\
    \heap(\addr).\mfree = 0
  }
  {{\tuple{\stack,\heap}},{\free(\var)} \step \tuple{\stack,\heap[\addr \mapsto \heap(\addr)[\mfree \mapsto 1]]}}
  \and
  \inferH{FreeAbort}
  { \stack(\var) = \addr\\
    \heap(\addr).\mfree \neq 0
  }
  {{\tuple{\stack,\heap}},{\free(\var)} \step \fail}
  
\end{mathpar}
\caption{Operational semantics of basic commands.\label{fig:concrete-semantics}}
\end{figure}

\begin{example}
  \Cref{fig:example-program} shows the pseudocode for a program $\prog$ that non-deterministically allocates a singly-linked list of arbitrary length and then deallocates the list again. The encoding of while loops into control-flow edges is as expected (e.g., there is an edge $\tuple{\loc_2,\assume(\varp = \pnull),\loc_3}$ to encode the loop exit transition for the second while loop).  The right side in \Cref{fig:example-program} depicts a state $\tuple{\ell_2,\tuple{\stack,\heap}} \in \den{\prog}$ with $\dom(\heap)=\set{\pnull,\addr_1,\dots,\addr_4}$, $\stack(\var)=\addr_1$, and $\stack(\varp)=\stack(\varpp)=\addr_3$. Freed nodes in $\heap$ are shaded in gray. We have $\den{P} \neq \top$, so the program is memory safe.
\end{example}


\section{View Abstraction}
\label{sec:view-abstraction}

We now formally define our \emph{view abstraction}, which is inspired by the idea of \emph{space invariants} in \cite{DBLP:conf/lpar/KahsaiKRS17}. A technical improvement over \cite{DBLP:conf/lpar/KahsaiKRS17} is that we define this abstraction in terms of the framework of abstract interpretation, which arguably leads to a cleaner design and makes the abstraction more easily reusable in combination with other abstractions. In fact, our formalization serves as an intermediate stepping stone towards the definition of our overall static analysis in \cref{sec:flow-abstraction}. That said, the view abstraction is also of independent interest as evidenced by \cite{DBLP:conf/lpar/KahsaiKRS17}.

At a high level, the idea of the view abstraction is as follows. An abstract state $\state^\viewdesignation$ only tracks those heap entries whose addresses are pointed to by some stack variable in $\state^\viewdesignation$. We refer to this part of the heap as the \emph{view heap}. By definition, the size of a view heap is statically bounded. The remainder of a (concrete) heap is abstracted by a so-called \emph{space invariant} $\hinv$. Intuitively, $\hinv$ is a set of tuples $\tuple{\addr,\fldval}$ such that $\heap(\addr)=\fldval$ for some state $\tuple{\loc,\tuple{\stack,\heap}} \in \den{\prog}$ where $\stack$ does not point to $\addr$. The space invariant is context-insensitive, i.e., it does not track the relationship between $\loc$ and $\heap$.

If during the abstract execution of a statement $\stmt$, a command reads from or writes to a field of an address $\addr$ that is not yet present in the view heap, then the abstract semantics nondeterministically chooses an entry $\tuple{\addr,\fldval} \in \hinv$ and adds it to the view heap. It then executes the command as in the concrete semantics. The extension of the view heap with entries drawn from $\hinv$ is referred to as a \emph{pull} in~\cite{DBLP:conf/lpar/KahsaiKRS17} and can be thought of as a form of \emph{materialization}~\cite{DBLP:journals/toplas/SagivRW02}. Conversely, if after execution of a statement a heap entry $\heap(\addr)$ is no longer pointed to by any stack variables, then $\heap(\addr)$ is evicted from the view heap and $\tuple{\addr,\heap(\addr)}$ is added to the invariant. This step is referred to as a \emph{push} in \cite{DBLP:conf/lpar/KahsaiKRS17}.

In the remainder of this section, we first define the abstract domain of the view abstraction and derive its best abstract transformer algebraically from the concrete transformer $\transformer$ using a Galois connection. In a second step, we then characterize this best abstract transformer operationally to recapture the intuition behind the analysis that we provided above.

\subsection{Abstract Domain and Transformer}

We define the set of \emph{view states} as $\viewstates = \locs \times \{\fail\} \cup \pset{\tuple{\loc,\tuple{\stack,\heap}}}{\dom(\heap) = \stack(\vars_p)}$. We refer to the heap component of a view state as a view heap. 
Note that view heaps $\heap$ are allowed to have pointers to addresses outside of $\dom(\heap)$ and are therefore not necessarily safe.

The abstract domain for the view abstraction is
$\viewdom = (\powersetof{\viewstates} \times \powersetof{\addrs \times \fldvals})_\fail$ with the partial order $\vieword$ defined as component-wise subset inclusion, extended with $\fail$. Thus, $\viewdom$ forms a complete lattice. We refer to an element $\view = \tuple{\stateset,\hinv} \in \viewdom$ as a \emph{view} and to $\hinv$ as the \emph{space invariant} of the view.

The \emph{completion} of a view heap $\heap$ subject to a space invariant $\hinv$, denoted $\hinv \hinvcomp \heap$, consists of the set of all safe heaps $\heapp$ that extend $\heap$ according to the entries in $\hinv$:
\[\hinv \hinvcomp \heap = \pset{\heapp}{\exists \heappp \in \heaps.\, \heappp \subseteq \hinv \land \heapp = \heap \uplus \heappp}\enspace.\]

The concretization function that relates $\viewdom$ with the concrete domain $\concdom$ is then defined by lifting the completion function to view states in the expected way:
\begin{align*}
\viewgamma(\stateset,\hinv) = \; & \pset{\tuple{\loc,\tuple{\stack,\heapp}}}{\tuple{\loc,\tuple{\stack,\heap}} \in \stateset \land \heapp \in \hinv \triangleright \heap} \quad \text{and} \quad \gamma(\fail) = \fail \enspace.
\end{align*}

We define a corresponding $\viewalpha \in \concdom \to \viewdom$ as the best abstraction function relative to $\viewgamma$:
\begin{align*}
\viewalpha(\stateset) = \bigviewmeet \pset{\view}{\stateset \concord \viewgamma(\view)}\enspace.
\end{align*}

\begin{restatable}{proposition}{viewGalois}
\label{prop:view-galois}
  $\tuple{\viewalpha,\viewgamma}$ form a Galois connection.
\end{restatable}

\begin{lemma}[Characterization of $\viewalpha$]
  \label{lem:view-alpha}
  For all $\stateset \in \powersetof{\states}$, we have
  \[\viewalpha(\stateset) = \bigviewjoin \pset{\tuple{\set{\tuple{\loc,\tuple{\stack,\restrict{\heap}{\rng(\stack)}}}},\restrict{\heap}{(\addrs \setminus \rng(\stack))}}}{\tuple{\loc,\tuple{\stack,\heap}} \in \stateset} \enspace.\]
  Moreover, $\viewalpha(\fail)=\fail$.
\end{lemma}

\begin{example}
  \label{ex:view-abstraction}
  Consider again the state $\state = \tuple{\loc_2,\tuple{\stack,\heap}}$ depicted on the right of \cref{fig:example-program}. Its abstraction $\viewalpha(\set{\state})$ is $\tuple{\set{\tuple{\loc_2,\tuple{\stack,\restrict{\heap}{\rng(\stack)}}}}, \restrict{\heap}{\addrs \setminus \rng(\stack)}}$ with
  \begin{align*}
    \restrict{\heap}{\rng(\stack)} = \{ & \addr_1 \mapsto \tuple{l: \addr_3,n: \pnull,\mfree:0}, \addr_3 \mapsto \tuple{l: \pnull, n: \addr_2,\mfree:0} \} \text{ and }\\
    \restrict{\heap}{\addrs \setminus \rng(\stack)} = \{ & \addr_4 \mapsto \tuple{l: \pnull, n: \addr_3,\mfree:1}, \addr_2 \mapsto \tuple{l: \pnull, n: \pnull,\mfree: 0},\\
   & \pnull \mapsto \tuple{l: \pnull, n: \pnull,\mfree:1}\} \enspace.
  \end{align*}
\end{example}

The \emph{view semantics} $\den{\prog}^\viewdesignation \in \viewdom$ of $\prog$ is obtained as the least fixpoint of the best abstract transformer with respect to the defined Galois connection:
\[\den{\prog}^\viewdesignation = \lfp{\viewtransformer} \quad \text{where} \quad
  \viewtransformer = \viewalpha \circ \transformer \circ \viewgamma\enspace.\]
Soundness of the view semantics follows by construction~\cite{DBLP:conf/popl/CousotC79,DBLP:conf/popl/CousotC77}.

\begin{corollary}
  For all programs $\prog$, $\den{\prog} \concord \viewgamma(\den{\prog}^\viewdesignation)$.
\end{corollary}

\subsection{Characterization of Best Abstract Transformer}
We next describe the best abstract transformer more operationally. To this end, we introduce an abstract transition relation that executes commands \emph{locally} in the view state, pulling heap entries from the space invariant $\hinv$ as needed.
Formally, we write $\stackheap, \com \viewstep_\hinv \stackheapp_\top$ to mean that a view state with stack and heap $\stackheap$ can take an abstract transition under command $\com$ and space invariant $\hinv$, to a stack-heap pair $\stackheapp$ (or fail with $\top$).

The definition of the abstract transition relation is shown in \cref{fig:view-semantics}. It is obtained calculationally from the best abstract transformer. We only show some of the more interesting rules. The remaining rules for the missing commands are directly obtained from the corresponding rules in \cref{fig:concrete-semantics}. We also omit the rules for the failure cases. As for the concrete transition relation, we extend this relation to an abstract transition relation on statements as expected.

\begin{figure}
  \def \MathparLineskip {\lineskip=1.3ex}
\begin{mathpar} 
    \inferH{VRead${}_d$}
    {\stack(\varp) = \addr\\
      \heap(\addr).\mfree = 0\\
      \heap(\addr).\fld_d = n
    }
    {{\tuple{\stack,\heap}},{\var := \varp.\fld_d} \viewstep_\hinv \tuple{\stack[\var \mapsto n],\heap}}
    \and
    \inferH{VRead${}_p$}
    {\stack(\varp) = \addr\\
      \heap(\addr).\mfree = 0\\
      \heap(\addr).\fld_p = \addrp\\
      \addrp \in \dom(\heap)\\
    }
    {{\tuple{\stack,\heap}},{\var := \varp.\fld_p} \viewstep_\hinv \tuple{\stack[\var \mapsto \addrp],\heap}}
    \and
    \inferH{VPull}
    {\stack(\varp) = \addr\\
      \heap(\addr).\mfree \!=\! 0\\
      \heap(\addr).\fld_p \!=\! \addrp\\
      \addrp \!\notin\! \dom(\heap)\\
      \tuple{\addrp,\fldval} \!\in\! \hinv
    }
    {{\tuple{\stack,\heap}},{\var := \varp.\fld_p} \viewstep_\hinv \tuple{\stack[\var \mapsto \addrp],\heap[\addrp \mapsto \fldval]}}
    \and
    \inferH{VWrite}
    { \eval{\expr}{\stack} = \val\\
      \stack(\var) = \addr\\
      \addr \in \dom(\heap)\\
      \heap(\addr).\mfree = 0
    }
    {{\tuple{\stack,\heap}},{\var.\fld := \expr} \viewstep_\hinv \tuple{\stack,\heap[\addr \mapsto \heap(\addr)[\fld \mapsto \val]]}}
    \and
    \inferH{VMalloc}
    {\addr \notin \dom(\heap)\\
      \fldval \in \initfldvals
    }
    {{\tuple{\stack,\heap}},{\var := \malloc} \viewstep_\hinv \tuple{\stack[\var \mapsto \addr],\heap[\addr \mapsto \fldval]}}
  \end{mathpar}
  
\caption{View semantics of basic commands. Some rules are omitted.\label{fig:view-semantics}}
\end{figure}

Note that when we have $\stackheap, \stmt \viewstep_\hinv \stackheapp$, then $\stackheapp$ is not necessarily a view state because the abstract transition relation does not evict entries from the heap component that are no longer directly pointed to by stack variables. To obtain a view state from $\stackheap$ we must propagate these entries back to the space invariant. To this end, we define the function
\[
  \push(\ell',\tuple{\stackp\,\heapp},\hinv) = \tuple{\set{\tuple{\ell',\tuple{\stackp,\restrict{\heapp}{\rng(\stackp)}}}}, \hinv \cup \restrict{\heapp}{\addrs \setminus \rng(\stackp)}}\enspace.
\]

We define the operational best abstract view transformer as
\begin{align*}
\viewpostof{\tuple{\loc,\stmt,\locp}}(\stateset,\hinv) = \; &
  \bigviewjoin \pset{\push(\locp,\stackheapp_\fail,\hinv)}{\exists \tuple{\loc,\stackheap} \in \stateset.\,
  \stackheap,\stmt \viewstep_\hinv \stackheapp_\fail} \sqcup {} \\
  & \bigviewjoin \pset{\fail}{\exists \tuple{\loc,\stackheap} \in \stateset.\,
  \stackheap,\stmt \viewstep_\hinv \fail}\enspace.
\end{align*}

\begin{restatable}{theorem}{viewLocalPost}
  \label{thm:view-local-post}
  $\den{\prog}^\viewdesignation = \lfp {\lambda \view.\, (\bigviewjoin_{\edge \in \edges} \viewpostof{\edge}(\view) \sqcup \tuple{\init,\emptyset})}$.
\end{restatable}

An analysis that utilizes view abstraction is useful for inferring quantified invariants about heaps that capture local properties expressing how the fields of an individual object relate to each other. However, the view abstraction is in general to imprecise for proving memory safety, as the following example demonstrates.

\begin{example}
  \label{ex:view-post}
  Consider again the program $\prog$ shown in \cref{fig:example-program}. We have $\den{\prog}^\viewdesignation = \fail$ even though $\den{\prog} \neq \fail$. That is, the view abstraction is too imprecise to prove memory safety of this program.

  To see that this is indeed the case,
  assume that $\den{\prog}^\viewdesignation = \tuple{\stateset,\hinv} \neq \fail$. Consider again the state $\state = \tuple{\loc_2,\tuple{\stack,\heap}}$ from \cref{ex:view-abstraction}. We have $\state \in \den{\prog}$. Since the semantics of \lstinline+malloc+ nondeterministically chooses an unused address, we also have $\statep \in \den{\prog}$ where $\statep=\tuple{\loc_2,\tuple{\stackp,\heapp}}$ is isomorphic to $\state$ except that $\addr_2$ and $\addr_4$ are interchanged. Because $\den{\prog} \subseteq \viewgamma{\den{\prog}^\viewdesignation}$, we must also have $\viewalpha(\set{\state,\statep}) \vieword \tuple{\stateset,\hinv}$. It follows that $\tuple{\addr_2,\tuple{l:\pnull,n:\addr_3,\mfree:1}} \in \hinv$ and $\tuple{\loc_2, \tuple{\stack,\restrict{\heap}{\rng(\stack)}}} \in \stateset$. Now consider the computation of the local abstract post on the view state $\tuple{\loc_2, \tuple{\stack,\restrict{\heap}{\rng(\stack)}}}$ for the statement $\stmt$ that corresponds to the body of the while loop at location $\loc_2$ of $\prog$. We have
  \[\tuple{\stack,\restrict{\heap}{\rng(\stack)}},\stmt \viewstep_\hinv \tuple{\stackp',\heapp'}\]
  where $\stackp'=\set{\var \mapsto \addr_0, \varp \mapsto \addr_2, \varpp \mapsto \addr_2}$ and
  \begin{align*}
  \heapp' = \{ & \addr_1 \mapsto \tuple{l: \addr_3,n: \pnull,\mfree:0}, \addr_3 \mapsto \tuple{l: \pnull, n: \addr_2,\mfree:1},
   \addr_2 \mapsto \tuple{l: \pnull, n: \addr_3,\mfree:1}\}\enspace.
  \end{align*}
  The reason is that when executing the first command \lstinline+z = y.n+, the rule \ruleref{VPull} may choose $\tuple{\addr_2,\tuple{l:\pnull,n:\addr_3,\mfree:1}} \in \hinv$. It follows that we must have $\tuple{\loc_2,\tuple{\stackp',\restrict{\heapp'}{\rng(\stackp')}}} \in \stateset$. However, we now have $\tuple{\stackp',\restrict{\heapp'}{\rng(\stackp')}},\stmt \viewstep_\hinv \fail$: \lstinline+z = y.n+ will aboard since $\heapp'(\addr_2).\mfree=1$. Thus, we conclude $\den{\prog}^\viewdesignation = \fail$.

  The challenge when proving memory safety of $\prog$ is that the analysis must capture the invariant that for all $\tuple{\loc_2,\tuple{\stack,\heap}} \in \den{\prog}$ if $\heap(\addr).\mfree=1$ for some address $\addr$ that is reachable from $\stack(\var)$, then $\addr = \pnull$. However, the view abstraction is too imprecise to capture such reachability information.
\end{example}



\section{Flow-based View Abstraction}
\label{sec:flow-abstraction}

To improve the precision of the view abstraction, we augment the concrete heap with an additional ghost field that captures non-local information about the heap (e.g., related to reachability).
To this end, we adapt the \emph{flow framework}~\cite{DBLP:journals/pacmpl/KrishnaSW18,DBLP:conf/esop/KrishnaSW20,DBLP:conf/tacas/MeyerWW23} for reasoning about inductive properties of heap graphs. Intuitively, the flow framework augments every node in the heap with a flow value according to a fixpoint equation that is reminiscent of a data flow analysis (except that it operates over the heap graph rather than the control flow graph of a program). The flow values are drawn from a \emph{flow domain}, which specifies how flow values are propagated through the graph and how they are aggregated at each node. By instantiating the framework with different flow domains, one can track different kinds of inductive properties, e.g., related to shape (such as whether a part of the heap forms a DAG, tree, or list) and functional correctness~\cite{DBLP:journals/pacmpl/KrishnaSW18,DBLP:conf/pldi/KrishnaPSW20}. 

The flow framework is designed so that it facilitates local reasoning about updates to the heap graph that affect the flow\footnote{In fact, the flow framework was originally conceived in the context of separation logic, but we will use it here at a purely semantic level without embedding it into a program logic.}. This feature is key to our design of a program analysis that builds on the view abstraction, yet, is able to infer space invariants that capture inductive heap properties.

\subsection{Flows and Flow Graphs}

\newcommand{\defFunc}[2]{\newcommand{#1}{\mathsf{#2}}}

\defFunc{\inflowfld}{in}
\defFunc{\outflowfld}{out}

A \emph{flow domain} $(\mdom, \mop, \mzero, \mord, \edgefn)$ consists of
an ordered commutative cancellative monoid $(\mdom, \mop, \mzero, \mord)$ and an \emph{edge function} $\edgefn \in \fldvals \to \addrs \to \mdom \to \mdom$. Intuitively, given a (partial) heap $\heap$, the edge function induces a labeled (partial) graph, which we call the \emph{flow graph} of $\heap$. The flow graph consists of nodes $\dom(\heap)$ and edges $\dom(\heap) \times \addrs$. Each edge $(\addr,\addrp)$ is labeled by the function $\edgefn(\heap(\addr))(\addrp) \in \mdom \to \mdom$. To reduce notational clutter, we write $\edgefn_\heap(\addr,\addrp)$ for $\edgefn(\heap(\addr))(\addrp)$ in the following. The edge label $\edgefn_\heap(\addr,\addrp)$ determines how flow values $\mval \in \mdom$ are routed from $\addr$ to $\addrp$. The case of $\edge_\heap(\addr,\addrp) = (\lambda \mval.\, \mzero)$ can, thus, be interpreted as describing the absence of an edge between $\addr$ and $\addrp$. We require that the presence of an edge must be supported by a corresponding physical link in the heap. That is, $\edgefn$ must satisfy that for all $\fldval$, if $\edgefn(\fldval)(\addrp) \neq (\lambda \mval.\, \mzero)$, then there exists $\fld \in \flds_p$ such that $\fldval.\fld=\addrp$. Note that the edge label only depends on the valuations of the fields at $\addr$ in $\heap$. 

\begin{example}
  \label{ex:flow-domain}
  The following flow domain can be used to compute information about reachability and sharing in heaps. Consider the natural numbers with addition as the monoid operation and the natural ordering, $(\NN, +, 0, \leq, \edgefn)$, where the edge function $\edgefn$ is defined as follows:
  \[ \edgefn(\fldval)(\addrp)(\mval) = \begin{cases}
      1 & \text{if } \fldval.l = \addrp \\
      \mval & \text{or else if } \fldval.n = \addrp\\
      0 & \text{otherwise} \enspace.
    \end{cases}
  \]
  The rationale for this definition is that a node pointed to by an $l$ field is the start of a list. Thus, an $l$ edge generates a constant flow $1$ to ``mark'' the head of the list. On the other hand, $n$ fields just propagate the flow in order to indicate reachability from the head node.
  For the heap $\heap$ depicted in \cref{fig:example-program}, we have $\edgefn_\heap(\addr_1,\addr_3) = (\lambda \mval.\, 1)$, $\edgefn_\heap(\addr_4,\addr_3) = (\lambda \mval.\, \mval)$, and $\edgefn_\heap(\addr_3,\addr_4) = (\lambda \mval.\, 0)$.  In the remainder of this section, we refer to this flow domain as the \emph{path-counting} flow domain.

\end{example}

The flow monoid $(\mdom, \mop, \mzero, \leq)$ and the flow graph define a data flow analysis that decorates each node in the graph with a flow value. These flow values are propagated along outgoing edges of each node according to the edge labels. The incoming flow values at each node are then aggregated using the monoid operation $\mop$ to compute the new flow value at that node. We formalize this process using the following \emph{flow transformer} $\flowtransformer{\heap,\inflow} \in (\addrs \to \mdom) \to \addrs \to \mdom$:
\[\flowtransformer{\heap,\inflow} = \lambda \flowvar.\, \lambda \addrp.\, \inflow(\addrp) \mop \kern-3pt \mBigOp_{\addr \in \dom(\heap)} \edgefn_\heap(\addr, \addrp)(\flowvar(\addr))\]
where $\inflow \in \addrs \to \mdom$ determines the initial \emph{inflow} into the flow graph. We require $\inflow(\addr) = \mzero$ for all $\addr \in \addrs \setminus \dom(\heap)$.

We refer to the restriction of $\lfp{\flowtransformer{\heap,\inflow}}$ to $\dom(\heap)$ as the \emph{flow} of $\heap$ with respect to inflow $\inflow$. Note that unless we impose additional assumptions on the flow domain or $\heap$, the least fixpoint of $\flowtransformer{\heap,\inflow}$ need not exist. We therefore define
\[
  \flow{\heap,\inflow} = \begin{cases}
    \restrict{\lfp{\flowtransformer{\heap,\inflow}}}{\dom(\heap)} & \text{if defined}\\
    \top & \text{otherwise}\enspace.
  \end{cases}
\]
As we shall see later, parameterizing the flow over an inflow is useful for reasoning locally about updates to the heap that affect its flow. However, when we focus our attention on the \emph{global} heap $\heap$ of a state $\tuple{\loc,\tuple{\stack,\heap}}$, then we always fix the inflow to the canonical \emph{zero inflow}, $\inflow_\mzero = (\lambda \addr.\, \mzero)$. In this case, we write just $\flow{\heap}$ for $\flow{\heap,\inflow_\mzero}$.

Similar to the inflow, we can also define the \emph{outflow} of a heap $\heap$ with respect to inflow $\inflow$. It is the function $\outflow{\heap,\inflow} \in \addrs \to \mdom$ defined as
\[
  \outflow{\heap,\inflow} = \begin{cases}
    \lambda \addr.\, \ite{\addr \notin \dom(\heap)}{\lfp{\flowtransformer{\heap,\inflow}}(\addr)}{\mzero} & \text{if defined}\\
    \top & \text{otherwise}\enspace.
  \end{cases}
\]

\begin{example}
  Consider again the path-counting flow domain of \cref{ex:flow-domain}. For the heap $\heap$ depicted in \cref{fig:example-program}, we have $\flow{\heap} = \{ \addr_1,\addr_4 \mapsto 0, \addr_2,\addr_3,\pnull \mapsto 1\}$. Intuitively, $\flow{\heap}(\addr) = \mval$ indicates that $\addr$ is reachable from $\mval$ root nodes in $\heap$ by following one \lstinline+l+ pointer and then an arbitrary number of \lstinline+n+ pointers. We also have $\outflow{\heap,\inflow_\mzero} = (\lambda \addr.\, \mzero)$. Let $\heapp$ be the heap obtained from $\heap$ by rerouting the outgoing \lstinline+n+ of $\addr_2$ to $\addr_4$, then $\flow(\heapp)=\top$. In general, for the path-counting flow domain, the flow only exists if there are no cycles among the paths labeled $\mathtt{l}\mathtt{n}^*$.
\end{example}

The following lemma states that we can compute the flow of a safe heap $\heap_1 \uplus \heap_2$ locally in each subheap $\heap_1$ and $\heap_2$. The corresponding inflows and outflows of the subheaps are dual to each other. Their uniqueness follows from the cancellativity of the flow monoid.

\begin{lemma}
  Let $\heap_1$ and $\heap_2$ be disjoint heaps such that $\heap = \heap_1 \uplus \heap_2$ is safe and $\flow(\heap) \neq \top$, then there exist unique inflows $\inflow_1$ and $\inflow_2$ such that $\flow(\heap_1,\inflow_1) = \restrict{\flow{\heap}}{\dom(\heap_1)}$, $\flow(\heap_2,\inflow_2) = \restrict{\flow(\heap)}{\dom(\heap_2)}$, $\outflow(\heap_1,\inflow_1) = \inflow_2$, and $\outflow(\heap_2,\inflow_2) = \inflow_1$.
\end{lemma}

\subsection{Abstract Domain and Transformer}

We define our flow-based view abstraction as a composition of two simpler abstractions. The first of these abstractions, which we refer to as the \emph{flow abstraction}, augments the heap $\heap$ in each state with a ghost field $\flowfld$ that stores the value $\flow(\heap)(\addr)$ at each $\addr \in \dom(\heap)$. The second abstraction is the view abstraction from \cref{sec:view-abstraction}, defined as before but over the augmented heaps. In the following, we make this abstraction formally precise.

We denote by $\flowheaps$ the set of heaps over the extended set of fields $\flds \uplus \set{\flowfld}$ where we require that valuations of $\flowfld$ take flow values from $\mdom$. Similarly, we denote by $\flowstates$ the set of states with heaps drawn from $\flowheaps$. We refer to the elements of $\flowheaps$ and $\flowstates$ as \emph{flow heaps} and \emph{flow states}, respectively. A flow heap $\heap$ is called \emph{consistent} if $\heap.\flowfld = \flow{\heap,\inflow}$ for some inflow $\inflow$. In this case, we denote this unique $\inflow$ by $\heap.\inflowfld$ and otherwise define $\heap.\inflowfld = \top$. Similarly, we denote $\outflow(\heap,\inflow)$ by $\heap.\outflowfld$ for consistent $\heap$ or else define $\heap.\outflowfld=\top$.

The abstract domain of the flow abstraction is the topped powerset lattice of flow states, $\flowdom = \powersetof{\flowstates}_\fail$. The concretization function $\flowgamma \in \flowdom \to \concdom$ simply filters out inconsistent flow states:
\[\flowgamma(\stateset^\flowdesignation) = \pset{\tuple{\loc,\tuple{\stack,\heap}}}{\flow(\heap) \neq \top \land \tuple{\loc,\tuple{\stack,\heap[\flowfld \mapsto \flow{\heap}]}} \in \stateset^\flowdesignation} \quad \text{and} \quad \flowgamma(\fail)=\fail\enspace.\]
The corresponding abstraction function $\flowalpha \in \concdom \to \flowdom$ augments each state in a set $\stateset$ with the flow component, provided flows exist for all states in $\stateset$, and otherwise sends $\stateset$ to $\fail$:
\begin{align*}
\flowalpha(\stateset) & {} = \pset{\tuple{\loc,\tuple{\stack,\heap[\flowfld \mapsto \flow(\heap)]}}}{\flow{\heap} \neq \top \land \tuple{\loc,\tuple{\stack,\heap}} \in \stateset} & \text{and} \quad \flowalpha(\fail) & {} = \fail \enspace.
\\
& \flowjoin \bigflowjoin \pset{\fail}{\exists \tuple{\loc,\tuple{\stack,\heap}} \in \stateset.\, \flow(\heap)=\top}\\
\end{align*}

A simple calculation shows that both $\flowgamma$ and $\flowalpha$ are monotone, $\flowgamma \circ \flowalpha$ is extensive, and $\flowalpha \circ \flowgamma$ is reductive. The following proposition then follows from \cite[Theorem 5.3.0.4]{DBLP:conf/popl/CousotC79}.
\begin{proposition}
  The pair $\tuple{\flowalpha,\flowgamma}$ forms a Galois connection.
\end{proposition}

We define $\flowviewalpha = \viewalpha \circ \flowalpha$ and $\flowviewgamma = \flowgamma \circ \viewgamma$. As Galois connections compose, $\tuple{\flowviewalpha,\flowviewgamma}$ forms a Galois connection between $\concdom$ and $\flowviewdom$, where $\flowviewdom$ is defined as $\viewdom$ in \cref{sec:view-abstraction}, except that $\flds^\flowdesignation = \flds \uplus \set{\flowfld}$ takes the role of $\flds$. 

The abstract semantics $\den{\prog}^\flowviewdesignation$ of a program $\prog$ with respect to the flow-based view abstraction is
\[\den{\prog}^\flowviewdesignation = \lfp{\flowviewtransformer} \quad \text{where} \quad
  \flowviewtransformer = \flowviewalpha \circ \transformer \circ \flowviewgamma\enspace.\]
Soundness of the flow-based view semantics again follows by construction.

\begin{corollary}
  \label{thm:flow-view-post-sound}
  For all programs $\prog$, $\den{\prog} \concord \flowviewgamma(\den{\prog}^\flowviewdesignation)$.
\end{corollary}

\begin{example}
  \label{ex:flow-view-post}
  Consider again the states $\state$ and $\statep$ from \cref{ex:view-post}. For the path-counting flow domain we have $\flowviewalpha(\set{\state,\statep}) = \tuple{\set{\tuple{\loc_2,\tuple{\stack,\restrict{\flowheap}{\rng(\stack)}}}, \tuple{\loc_2,\tuple{\stackp,\restrict{\flowheapp}{\rng(\stackp)}}}}, \hinv}$ with
  \begin{align*}
    \restrict{\flowheap}{\rng(\stack)} = \{ & \addr_1 \mapsto \tuple{l: \addr_3,n: \pnull,\mfree:0,\flowfld:0}, \addr_3 \mapsto \tuple{l: \pnull, n: \addr_2,\mfree:0,\flowfld: 1} \}\\
    \restrict{\flowheapp}{\rng(\stackp)} = \{ & \addr_1 \mapsto \tuple{l: \addr_3,n: \pnull,\mfree:0,\flowfld:0}, \addr_3 \mapsto \tuple{l: \pnull, n: \addr_4,\mfree:0,\flowfld: 1} \} \\
    \hinv = \{ & \tuple{\addr_4, \tuple{l: \pnull, n: \addr_3,\mfree:1, \flowfld: 0}}, \tuple{\addr_2, \tuple{l: \pnull, n: \pnull,\mfree: 0, \flowfld: 1}}, \\
& \tuple{\addr_2, \tuple{l: \pnull, n: \addr_3,\mfree:1, \flowfld: 0}}, \tuple{\addr_4, \tuple{l: \pnull, n: \pnull,\mfree: 0, \flowfld: 1}},\\
& \pnull \mapsto \tuple{l: \pnull, n: \pnull,\mfree:1, \flowfld: 1}\} \enspace.
  \end{align*}
  Similar to the precision loss under view abstraction in \cref{ex:view-post}, the set $\viewgamma(\flowviewalpha(\set{\state,\statep}))$ contains flow states with flow heaps such as this one:
  \[\begin{array}{l} \{ \addr_1 \mapsto \tuple{l: \addr_3,n: \pnull,\mfree:0,\flowfld:0},
    \addr_3 \mapsto \tuple{l: \pnull, n: \addr_2,\mfree:0,\flowfld: 1},\\
    \;\, \addr_2 \mapsto \tuple{l: \pnull, n: \addr_3,\mfree:1, \flowfld: 0},
    \pnull \mapsto \tuple{l: \pnull, n: \pnull,\mfree:1, \flowfld: 1}\}\enspace.
    \end{array}
  \]
  However, these flow heaps are inconsistent and filtered out by $\flowgamma$. In fact, we have $\flowviewgamma(\flowviewalpha(\set{\state,\statep})) = \set{\state,\statep}$. The abstract domain $\flowviewdom$ is precise enough to capture the invariant needed for proving memory safety of $\prog$. That is, we obtain $\den{\prog}^\flowviewdesignation \neq \fail$.
\end{example}

\subsection{Local Abstract Transformer}

As in the case of the simple view abstraction, we would ideally like to characterize the abstract transformer $\flowviewtransformer$ in terms of an abstract transition relation that operates \emph{locally} on the view heap $\heap_0$ of a view state $\tuple{\stack,\heap_0}$, i.e., without ever explicitly materializing any complete heap in $\hinv \hinvcomp \heap_0$. Unfortunately, for the flow-based view abstraction, such a local abstract transformer is inherently elusive. A field update at an address $\addr$ within $\heap_0$ may change the edge label $\edgefn_{\heap_0}(\addr,\addrp)$ for any other address $\addrp$. In turn, the flow transformer may propagate this change arbitrarily far beyond the view heap into the unmaterialized part $\heap_1$ of the global heap that is abstracted by $\hinv$. It therefore seems inevitable that in order to compute $\flowviewtransformer$ precisely, one has to reason non-locally about the effects of updates in $\heap_0$ on the flow of the unmaterialized heaps $\heap_1$ abstracted by $\hinv$.

We therefore pursue a more modest goal: we want to capture the cases when the effect of an update of $\heap_0$ to $\heap'$ \emph{does not} spill into the unmaterialized $\heap_1$. In this case, we refer to the update as \emph{view-local}. For view-local updates, we will provide a means to effectively compute the abstract transformer in a local manner. In the remaining cases, our local abstract transformer will fail, safely approximating $\flowviewtransformer$. The rationale for this design is that, in practice, the view heap $\heap_0$ is usually sufficiently large to contain the flow update, thus yielding a transformer that is both efficiently implementable and sufficiently precise. 

The following condition adapted from \cite{DBLP:conf/tacas/MeyerWW23} characterizes view-local updates. Intuitively, the condition states that $\heap_0$ and $\heapp$ are contextually equivalent with respect to how flow is routed through the updated region:
\[
\synccond{\heap_0,\heapp} \Leftrightarrow (\forall \inflow.\, \inflow \mord \heap_0.\inflowfld \Rightarrow \outflow(\heap_0,\inflow) = \outflow(\heapp,\inflow))\enspace.
\]
Note that the condition allows $\heap_0$ and $\heapp$ to differ in their domains as long as the inflow and outflow to the nodes that are not included in both flow heaps is $\mzero$. This accounts for updates that involve allocation. The quantification over all inflows $\inflow \mord \heap_0.\inflowfld$ ensures that the outflow is preserved for all iterates of flows in larger flow graphs, not just their least fixpoint. We elide this technicality here as we will later discuss a simplification in \cref{sec:sync-optimization} that will eliminate this quantification.
The following lemma then states that $\synccond{\heap_0,\heapp}$ guarantees that the flow does not change in the context of the update.

\begin{lemma}
  \label{lem:frame-preserving-update}
  For consistent flow heaps $\heap_0 \uplus \heap_1$ and $\heapp$, if $\dom(\heapp) \cap \dom(\heap_1) = \emptyset$ and $\synccond{\heap_0,\heapp}$ then $\heapp \uplus \heap_1$ is consistent, $\heapp.\flowfld = \restrict{\flow{\heapp \uplus \heap_1}}{\dom(\heapp)}$, and $\heap_1.\flowfld = \restrict{\flow{\heapp \uplus \heap_1}}{\dom(\heap_1)}$. 
\end{lemma}


\begin{figure}
  \begin{minipage}{.5\textwidth}
  \centering
  \begin{tikzpicture}[>=stealth, font=\footnotesize, scale=0.8, every node/.style={scale=0.8}]
    \def\xsep{6}
    \def\ysep{6}

    \draw[draw=black] (-1.5,-.72) rectangle ++(2.6,3);
    \node at (.8,2) {$\heap_0$};

    \node[unode] (n3) {$\addr_3$};
    \node[unode, right=12 mm of n3] (n2) {$\addr_2$};
    \node[mnode, right=\xsep mm of n2] (null) {$\pnull$};
    \node[unode, above=\ysep mm of n3] (n1) {$\addr_1$};

    \node[stackVar, left=.4cm of n1] (x) {$\var$};
    \node[stackVar, left=.4cm of n3] (y) {$\varp$};
    \node[stackVar, above=.4cm of n2] (z) {$\varpp$};

    \draw[edge] (x) to (n1);
    \draw[edge] (y) to (n3);
    \draw[edge] (z) to (n2);
    \draw[edge] (n1) to node[left] {$\mathtt{l}$} (n3);
    \draw[edge] (n1) to node[right] {\textbf{1}} (n3);
    \draw[edge] (n3) to node[above,pos=.64] {$\mathtt{n}$} (n2);
    \draw[edge] (n3) to node[below,pos=.64] {\textbf{1}} (n2);
    \draw[edge] (n2) to node[above] {$\mathtt{n}$} (null);
    \draw[edge] (n2) to node[below] {\textbf{1}} (null);
  \end{tikzpicture}
  \end{minipage}%
  \hfill
  \begin{minipage}{.5\textwidth}
  \centering
  \begin{tikzpicture}[>=stealth, font=\footnotesize, scale=0.8, every node/.style={scale=0.8}]
    \def\xsep{6}
    \def\ysep{6}

    \draw[draw=black] (-1.5,-.72) rectangle ++(2.6,3);
    \node at (.8,2) {$\heap'$};

    \node[unode] (n3) {$\addr_3$};
    \node[unode, right=12 mm of n3] (n2) {$\addr_2$};
    \node[mnode, right=\xsep mm of n2] (null) {$\pnull$};
    \node[unode, above=\ysep mm of n3] (n1) {$\addr_1$};

    \node[stackVar, left=.4cm of n1] (x) {$\var$};
    \node[stackVar, left=.4cm of n3] (y) {$\varp$};
    \node[stackVar, above=.4cm of n2] (z) {$\varpp$};

    \draw[edge] (x) to (n1);
    \draw[edge] (y) to (n3);
    \draw[edge] (z) to (n2);
    \draw[edge] (n1) to node[above,pos=.57] {$\mathtt{l}$} (n2);
    \draw[edge] (n1) to node[below,pos=.57] {\textbf{1}} (n2);
    \draw[edge] (n3) to node[above,pos=.64] {$\mathtt{n}$} (n2);
    \draw[edge] (n3) to node[below,pos=.64] {\textbf{0}} (n2);
    \draw[edge] (n2) to node[above] {$\mathtt{n}$} (null);
    \draw[edge] (n2) to node[below] {\textbf{1}} (null);
  \end{tikzpicture}    
  \end{minipage}
  \caption{Illustration of updates following the scheme outlined in \cref{lem:frame-preserving-update}. The heap $\heap_0$ (left) is updated to $\heap'$ (right) without affecting the interface: node $\addr_2$ receives a combined flow value of $1$ before and after the update and the locality condition for the update is satisfied.\label{fig:interface-lemma}}
\end{figure}
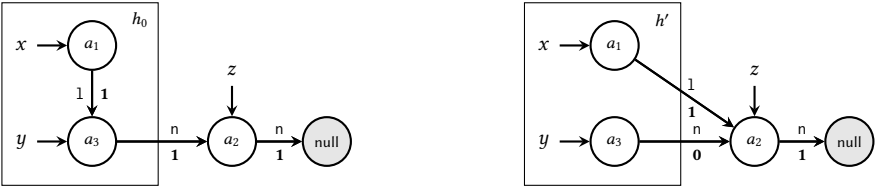

\Cref{fig:interface-lemma} visualizes the lemma.
The heaps $\heap_0$ (left of \cref{fig:interface-lemma}) and $\heap'$ (right of \cref{fig:interface-lemma}) produce the same outflow of $1$ to $\addr_2$.
It is easy to see that the locality condition $\synccond{\heap_0,\heapp}$ is satisfied.
Note that $\heap_0$ has no inflow which is why the universal quantifier in the condition is easy to discharge.

The idea for our local flow-based abstract transformer $\flowviewpost$ is then as follows. Like $\viewpost$ from \cref{sec:view-abstraction}, it executes a statement $\stmt$ locally starting from a view flow state $\tuple{\stack,\heap}$, materializing entries from $\hinv$ as needed. The execution produces an updated view flow state $\tuple{\stackp,\heapp}$ along with $\heap_0$, which is $\heap$ extended with all the entries pulled from $\hinv$. The abstract post operator then updates $\heapp.\flowfld$ to the new local flow of $\heapp$ and pushes the resulting view flow state back to $\hinv$, provided $\synccond{\heap_0,\heapp}$ holds:
\begin{align*}
\flowviewpostof{\tuple{\loc,\stmt,\locp}}(\stateset,\hinv) = \; &
  \bigflowviewjoin \pset{\sync(\locp,\stackheapp,\heap_0,\hinv)}{\exists \tuple{\loc,\tuple{\stack,\heap}} \in \stateset.\,
  \tuple{\stack,\heap},\heap,\stmt \flowviewstep_\hinv \stackheapp,\heap_0} \flowviewjoin {}\\
  & \bigflowviewjoin \pset{\fail}{\exists \tuple{\loc,\tuple{\stack,\heap}} \in \stateset.\,
  \tuple{\stack,\heap},\heap,\stmt \flowviewstep_\hinv \fail}
\end{align*}
where
\[
  \sync(\ell',\tuple{\stackp,\heapp},\heap_0,\hinv) = \begin{cases}
    \tuple{\emptyset, \emptyset} & \text{if } \heap_0.\inflowfld = \top\\
    \push(\ell',\tuple{\stackp,\heapp[\flowfld \mapsto \flow{\heapp,\heap_0.\inflow}]},\hinv) & \begin{aligned}[t]\text{if } &\heap_0.\inflowfld \neq \top \text{ and } \\ &\synccond{\heap_0,\heapp}\end{aligned} \\
    \fail & \text{otherwise}\enspace.
  \end{cases}
\]
The transition relation $\tuple{\stack,\heap},\heap,\stmt \flowviewstep_\hinv \stackheapp,\heap_0$ defining $\flowviewpost$ is shown in \cref{fig:flow-view-semantics}. We \new{highlight} the changes relative to the transition relation of the view semantics shown in \cref{fig:view-semantics}. Most of the rules are similar to the rules of the view semantics and just thread the additional component $\heap_0$ capturing the initial view heap through the transition. The most interesting rules are those related to reads from pointer fields whose target node $\addrp$ is not yet present in the view heap (\ruleref{FVPull} and \ruleref{FVPullAbort}) as well as allocation (\ruleref{FVMalloc}).

The rule \ruleref{FVPull} is the pendant to the rule \ruleref{VPull} for the view abstraction. It pulls an entry $\tuple{\addrp,\fldval}$ from the space invariant $\hinv$ and adds it to both the current view flow heap $\heap$ and the initial heap $\heap_0$. The condition $\heap.\outflowfld(\addrp) \leq \fldval.\flowfld$ reflects the improved precision compared to the earlier \ruleref{VPull} due to the tracking of the flow: the abstract transformer can disregard entries for $\addrp$ in $\hinv$ that do not respect the lower bound on the flow induced by $\heap$. The condition $\heap_0 = \restrict{\heap}{\dom(\heap_0)}$ ensures that no observable field writes have happened since the beginning of the execution of the current statement $\stmt$. This condition is needed for the soundness of the abstract transformer: if \ruleref{FVPull} were used after an observable field write, then $\synccond{\heap,\heap_0}$ may be (temporarily) violated. In this case, the flow value in the entries for $\addrp$ in $\hinv$ may no longer reflect the actual flow in the unmaterialized part of the global heap. We would have to materialize the global heap and recompute its flow in all such cases, going against the idea of a local abstract transformer. Therefore, the abstract transformer fails if $\heap_0 \neq \restrict{\heap}{\dom(\heap_0)}$ using rule \ruleref{FVPullAbort}.

The rule \ruleref{FVMalloc} handles allocation. It strengthens rule \ruleref{VMalloc} by requiring $\heap.\outflowfld(\addr)=\mzero$ for the freshly allocated address $\addr$. Since freed addresses are not reused, heaps are closed, and edges can only exist in the flow graph if they also exist in the heap, the flow of freshly allocated addresses must always be $\mzero$. Thus, if $\heap.\outflowfld(\addr)\neq \mzero$, then $\addr$ must already have been allocated before in the unmaterialized part of the global heap. The set $\initfldvals^\flowdesignation$ used in the last premise of the rule is defined like $\initfldvals$ except that it consists of all field valuations over $\flds^\flowdesignation$ whose ghost field $\flowfld$ is additionally initialized to $\mzero$.

The soundness proof of the local abstract transformer follows from these considerations about the abstract transition relation as well as \cref{lem:frame-preserving-update}.

\begin{figure}
  \def \MathparLineskip {\lineskip=1.3ex}
  \begin{mathpar} 
    \inferH{FVRead${}_d$}
    {\stack(\varp) = \addr\\
      \heap(\addr).\mfree = 0\\
      \heap(\addr).\fld_d = n
    }
    {{\tuple{\stack,\heap}},\new{\heap_0},{\var := \varp.\fld_d} \flowviewstep_\hinv \tuple{\stack[\var \mapsto n],\heap},\new{\heap_0}}
    \and
    \inferH{FVRead${}_p$}
    {\stack(\varp) = \addr\\
      \heap(\addr).\mfree = 0\\
      \heap(\addr).\fld_p = \addrp\\
      \addrp \in \dom(\heap)\\
    }
    {{\tuple{\stack,\heap}},\new{\heap_0},{\var := \varp.\fld_p} \flowviewstep_\hinv \tuple{\stack[\var \mapsto \addrp],\heap},\new{\heap_0}}
    \and
    \inferH{FVPull}
    {\stack(\varp) = \addr\\
      \heap(\addr).\mfree \!=\! 0\\
      \heap(\addr).\fld_p \!=\! \addrp\\
      \addrp \!\notin\! \dom(\heap)\\
      \tuple{\addrp,\fldval} \!\in\! \hinv\\
      \new{\heap_0 = \restrict{\heap}{\dom(\heap_0)}}\\
      \new{\heap.\outflowfld \neq \top}\\
      \new{\heap.\outflowfld(\addrp) \mord \fldval.\flowfld}
    }
    {{\tuple{\stack,\heap}},\new{\heap_0},{\var := \varp.\fld_p} \flowviewstep_\hinv \tuple{\stack[\var \mapsto \addrp],\heap[\addrp \mapsto \fldval]},\new{\heap_0[\addrp \mapsto \fldval]}}
    \and
    \inferH{FVPullAbort}
    {\stack(\varp) = \addr\\
      \heap(\addr).\mfree \!=\! 0 \\
      \heap(\addr).\fld_p \!=\! \addrp\\
      \addrp \!\notin\! \dom(\heap)\\
      \new{\heap_0 \neq \restrict{\heap}{\dom(\heap_0)}}
    }
    {{\tuple{\stack,\heap}},\new{\heap_0},{\var := \varp.\fld_p} \flowviewstep_\hinv \fail}
    \and
    \inferH{VWrite}
    { \eval{\expr}{\stack} = \val\\
      \stack(\var) = \addr\\
      \addr \in \dom(\heap)\\
      \heap(\addr).\mfree = 0
    }
    {{\tuple{\stack,\heap}},\new{\heap_0},{\var.\fld := \expr} \flowviewstep_\hinv \tuple{\stack,\heap[\addr \mapsto \heap(\addr)[\fld \mapsto \val]]},\new{\heap_0}}
    \and
    \inferH{FVMalloc}
    {\addr \notin \dom(\heap)\\
      \new{\heap.\outflow \neq \top}\\
      \new{\heap.\outflow(\addr) = \mzero}\\
      \fldval \in \new{\initfldvals^\flowdesignation}
    }
    {{\tuple{\stack,\heap}},\new{\heap_0},{\var := \malloc} \flowviewstep_\hinv \tuple{\stack[\var \mapsto \addr],\heap[\addr \mapsto \fldval]},\new{\heap_0}}
  \end{mathpar}
  
\caption{Local flow-based view semantics of basic commands. Some rules are omitted.\label{fig:flow-view-semantics}}
\end{figure}

\begin{theorem}
  \label{thm:local-flow-view-post-sound}
  $\den{\prog}^\flowviewdesignation \flowvieword \lfp \view.\, (\bigflowviewjoin_{\edge \in \edges} \flowviewpostof{\edge}(\view) \flowviewjoin \tuple{\init,\emptyset})$.
\end{theorem}

The overall soundness of the analysis is then implied by \cref{thm:flow-view-post-sound} and \cref{thm:local-flow-view-post-sound}.

\begin{example}
  Consider the abstract state $\tuple{\loc_2,\tuple{\stack,\restrict{\flowheap}{\rng(\stack)}}}$ and space invariant $\hinv$ from \cref{ex:flow-view-post}. Shortening $\restrict{\flowheap}{\rng(\stack)}$ to just $\heap$ in the following, we have
  $\tuple{\stack,\heap}, \heap, \stmt \flowviewstep_\hinv \tuple{\stackp,\heapp},\heap_0$ where
  \begin{align*}
    \stackp = \{ & \var \mapsto \addr_0, \varp,\varpp \mapsto \addr_2\}\\
    \heap_0 = \{ & \addr_1 \mapsto \tuple{l: \addr_3,n: \pnull,\mfree:0,\flowfld:0}, \addr_3 \mapsto \tuple{l: \pnull, n: \addr_2,\mfree:0,\flowfld: 1}, \\
   & \addr_2 \mapsto \tuple{l: \pnull, n: \pnull,\mfree:0,\flowfld: 1}\}\\
    \heapp = \{ & \addr_1 \mapsto \tuple{l: \new{\addr_2},n: \pnull,\mfree:0,\flowfld:0}, \addr_3 \mapsto \tuple{l: \pnull, n: \addr_2,\mfree:\new{1},\flowfld: 1}, \\
   & \addr_2 \mapsto \tuple{l: \pnull, n: \pnull,\mfree:0,\flowfld: 1}\}
  \end{align*}
\end{example}
Note that when the command \lstinline+z := y.n+ of $\stmt$ is executed, then rule \ruleref{FVPull} is applied for $\addrp = \addr_2$. The premise $\heap.\outflowfld(\addr_2) \mord \fldval.\flowfld$ of the rule ensures that the entry $\tuple{\addr_2,\fldval} \in \hinv$ with $\fldval.\flowfld = 1$ is pulled. We have $\synccond(\heap_0,\heapp)$. Thus, the call to $\sync(\ell_2,\tuple{\stackp,\heapp},\heap_0,\hinv)$ will update the flow of $\addr_3$ in $\heapp$ to $0$, and then evict $\addr_3$ from the updated view heap and push it to $\hinv$.



\subsection{A Light-weight $\mathit{ViewLocal}$ Condition}
\label{sec:sync-optimization}

View-local updates allow us to derive an easy-to-use and easy-to-implement transition system that replaces the abstract transformer.
However, view-local updates leave us with a hard task still: establishing the locality condition $\synccond{\bullet}$ involves quantification over certain inflows, making it difficult to check.
To address this, we propose a more lightweight alternative by further restricting the types of updates we are willing to support.
Specifically, we impose an acyclicity requirement on heap graphs, which eliminates the need for the quantifier in $\synccond{\bullet}$ and simplifies the entire $\sync$ operator.

In the setting of the flow framework, we say that a heap $\heap$ is acyclic if its induced flow graph is acyclic; more precisely, if the edges $\edgefn_\heap$ are acyclic.
As alluded to earlier, we rely on the understanding that the absence of flow, $\edge_\heap(\addr,\addrp) = (\lambda \mval.\, \mzero)$, between two nodes $\addr$ and $\addrp$ is interpreted as the absence of an edge between them.
Note that this means that a heap $\heap$ is acyclic by our definition, even if the actual object graph represented by $\heap$ would not traditionally be considered acyclic.
For instance, if the list from our example in \Cref{sec:motivation} was doubly-linked, we would not propagate flow via the back-edges, and thus the heap would be considered acyclic.

We now develop a new version of $\sync$ that leverages acyclicity.
It is easy to see that acyclicity guarantees the existence of the fixpoint of the flow transformer.
Moreover, the fixpoint can be computed effectively in a finite number of steps.
Intuitively, the fixpoint degenerates to a sum of all simple paths through the heap graph.

In order to rely on acyclicity in the presence of updates to the heap graph, we must ensure that acyclicity is maintained.
It is easy to ensure that the footprint remains acyclic.
However, the overall heap graph may become cyclic despite the footprint being acyclic.
In the example from \cref{sec:motivation}, connecting $\addr_2$ to $\addr_1$ makes the list form a ring while a footprint containing only $\addr_2$ and $\addr_1$ would still be acyclic.
What complicates the acyclicity check is that, as before, we wish not to materialize the heap graph beyond the local view.
To that end, we check a stronger property.
We enforce that the nodes in the footprint cannot reach nodes outside the footprint after the update unless they could already reach them before the update, again ignoring edges that do not carry flow.
To be more specific, let $\addr$ and $\addrp$ be nodes within and outside of the footprint, respectively.
The property we check is this: if there is a path from $\addr$ to $\addrp$ after the update, then there must be a path from $\addr$ to $\addrp$ before the update.
The existence of a path before the update guarantees, by acyclicity, that there is no path from $\addrp$ to $\addr$.
Hence, the update does not close a cycle.
We do not check this property for newly allocated nodes as they are guaranteed to not introduce cycles.
We use $\gacyc(\heapp)$ to mean that both the footprint is acyclic and the above property approximating acyclicity of the non-materialized graph holds.

More surprising is the interplay between acyclicity and the locality condition $\synccond{\bullet}$.
The requirement that all iterates of the footprint's flow before and after the update must produce the same outflow becomes obsolete.
It now suffices that the outflow for the actual flow of the footprint coincides before and after the update, effectively avoiding the quantifier over inflows.
To make this precise, we define a new, simplified locality condition: \[
	\synccondac{\heap_0,\heapp} \Leftrightarrow \outflow(\heap_0,.\heap_0\inflow) = \outflow(\heapp,\heap_0.\inflow) \land \gacyc(\heapp)
	\enspace.
\]
To get an intuitive understanding of this condition, assume that there is no path from a node within the footprint to another node within the footprint.
We argue that the flow values at all nodes \emph{succeeding} the footprint, i.e., to which a path from the footprint exists, is the same before and after the update.
To that end, consider how the flow for the graph is computed.
The fixpoint computation propagates (intermediate) flow values to the footprint.
These values are the inflow to the footprint and are then propagated through the footprint.
Without affecting the overall result of the flow computation, we can accelerate the fixed point in the region that has already received flow (the footprint and its \emph{preceding} nodes) without further propagating the flow to the nodes succeeding the footprint.
After a finite number of such accelerating steps, the footprint will obtain the flow values that the actual, unmodified fixed point would assign as its final result.
This is the case because our assumption guarantees that the footprint is not affected by its own outflow.
Now, $\synccondac$ states that outflow of the footprint after the update is the same as before the update.
This means that from that moment on, the flow computation for the succeeding nodes of the heap graph is the same before and after the update.
Hence, the flow at the succeeding nodes is the same before and after the update.

Technically, the soundness of $\synccondac$ relies on an application of Beki\'c's lemma~\cite{DBLP:conf/ibm/Bekic84e} to implement the acceleration of the flow fixpoint in the footprint without making assumptions to the graph beyond acyclicity.
The following lemma is the analogue of \Cref{lem:frame-preserving-update} for $\synccondac$.

\begin{lemma}
  For consistent acyclic flow heaps $\heap_0 \uplus \heap_1$ and $\heapp$, if $\dom(\heapp) \cap \dom(\heap_1) = \emptyset$ and $\synccondac{\heap_0,\heapp}$ then $\heapp \uplus \heap_1$ is consistent and acyclic, $\heapp.\flowfld = \restrict{\flow{\heapp \uplus \heap_1}}{\dom(\heapp)}$, and $\heap_1.\flowfld = \restrict{\flow{\heapp \uplus \heap_1}}{\dom(\heap_1)}$.
\end{lemma}



\section{Implementation}
\label{sec:impl}

We have implemented our proposed analysis in a tool called \atool{triceratops}. \atool{triceratops} accepts programs in a C-like input language, performing an \emph{instrumentation} by applying the flow-specific view abstraction from \Cref{sec:flow-abstraction} to produce a new \texttt{C} program that eliminates all heap usage.
During instrumentation, additional assertions are inserted to ensure memory safety.
The transformed program is then analyzed by an off-the-shelf solver to verify the correctness of these assertions.
We utilize a combination of the verification tool~\atool{tricera} \cite{DBLP:conf/fmcad/EsenR22} and the Horn solver~\atool{eldarica} \cite{DBLP:conf/fmcad/HojjatR18,DBLP:conf/fm/HojjatKGIKR12} for this.

Our instrumentation relies on several constructs and operators not native to \texttt{C} but widely supported by verification tools: (i)~\code{havoc} for non-deterministically choosing values (ii)~\code{assert} for raising a runtime error if a given condition fails, and (iii)~\code{assume} for blocking the execution if the given condition is not satisfied~\cite{DBLP:conf/popl/FlanaganS01}.
Additionally, our encoding of space invariants depends on (iv)~uninterpreted predicates for representing~$\aninv$, which are less commonly supported. Our back-end solvers, \atool{tricera} and \atool{eldarica}, support uninterpreted predicates.
While we are not aware of a general method to encode them in any verification tool, we believe it is straightforward to do so in Horn clause-based tools.

In the following, we elaborate on the inner workings of \atool{triceratops}.
\Cref{sec:impl:overview} gives an overview of the instrumentation and \cref{sec:impl:view-heaps,sec:impl:mem-safety,sec:impl:no-realloc,sec:impl:concurrency} go over its most notable specifics.

\subsection{Overview of the Instrumentation}
\label{sec:impl:overview}

The instrumentation performed by \atool{triceratops} rewrites the input program into a new program without heap usage, closely following the structure of the original program and retaining all control structures like loops and conditionals.
For primitive commands, it adheres to the discussion from \Cref{sec:flow-abstraction}, specifically the view semantics from \Cref{fig:flow-view-semantics}.
The three main ingredients for handling commands are:
\begin{inparaenum}
     \item using the view heap instead of the actual heap (related to all rules from \Cref{fig:flow-view-semantics}),
     \item pulling new entries into the view heap (rules \ruleref{FVPull} and \ruleref{FVPullAbort}), and
     \item pushing entries out of the view heap ($\push$ operator used in $\sync$ for defining $\flowviewpost$).
\end{inparaenum}
Below, we review how these ingredients are implemented in the instrumentation.

\paragraph{View Heap}
The original program accesses the heap by dereferencing pointers, such as \code{x.f} to access field \code{f} of the address $\addr$ that \code{x} points to.
These accesses are replaced to refer to the corresponding value from the view heap instead.
Given a view heap $\heap$, the expression \code{x.f} is replaced by $\heap(\addr).\mcode{f}$ from the view heap.
This replacement strategy is applied to all expressions in the input program.
The presence of the necessary entries in the view heap is ensured by our instrumentation for pulling.
For the exact encoding of $\heap$, refer to \Cref{sec:impl:view-heaps}.
While it mimics our technical development, we avoid using a dynamic map and instead encode $\heap$ in the stack.
Note that all pointer-type values in the original program become \code{int} values in the instrumented program.

\begin{figure}
    \begin{lstlisting}[mathescape=true,belowskip=0pt]
assert(x != NULL);                                                                 $\label[line]{code:ipull:null}$
$\heap$(x).field$_1$ = havoc; /* ... */ $\heap$(x).field$_k$ = havoc;              $\label[line]{code:ipull:havocm}$
$\heap$(x).$\mfree$ = havoc; $\heap$(x).$\flowfld$ = havoc;                        $\label[line]{code:ipull:havocs}$
assume($\aninv$(x, $\heap$(x).field$_1$, /* ... */, $\heap$(x).field$_k$, $\heap$(x).$\mfree$, $\heap$(x).$\flowfld$));      $\label[line]{code:ipull:inv}$
assert($\heap$(x).$\mfree$ == 0);                                                  $\label[line]{code:ipull:free}$
$\heap_0$(x) = $\heap$(x);                                                         $\label[line]{code:ipull:save}$
\end{lstlisting}
    \caption{An idealized instrumentation template for pulling pointer \code{x}. The view heap for \code{x} is materialized from the space invariant, and the pointer is checked to be safe to use. The template is idealized in that the view heap is not resolved down to our actual encoding (cf. \cref{sec:impl:view-heaps}).\label{fig:impl:pull}}
\end{figure}

\paragraph{Pulls}
When an expression like \code{x.f} requires an entry from the view heap $\heap$ for an address $\addr'$ pointed to by \code{x}, we extend $\heap$ to materialize entries from the space invariant for $\addr'$ if needed.
We refer to the program locations where this materialization happens as \emph{pull sites}.
We identify pull sites statically using a standard dataflow analysis on the original program to identify the pointers \code{x} that need pulling.
This analysis determines the earliest program location where \code{x} will definitely be used, minimizing the number of pull sites and preventing repeated pulls of the same address.

\Cref{fig:impl:pull} gives a template for pulling a pointer \code{x}.
As shown in \Cref{fig:flow-view-semantics}, pull sites materialize values for all fields of the pulled address $\addr'$.
This is encoded by non-deterministically choosing field values (\cref{code:ipull:havocm,code:ipull:havocs}) and then assuming that the choice is allowed by the space invariant (\cref{code:ipull:inv}).
Additionally, we ensure that \code{x} is neither $\pnull$ (\cref{code:ipull:null}) nor freed (\cref{code:ipull:free}) for subsequent dereferences to be safe (cf. \cref{sec:impl:mem-safety}).
Finally, we create a backup $\heap_0(\mcode{x})$ of the materialized fields $\heap(\mcode{x})$ (\cref{code:ipull:save}) to reconcile the flow interface before and after updates (cf. \Cref{sec:view-abstraction,fig:flow-view-semantics}).

Finally, we safely over-approximate the premise $\heap.\outflowfld(\addr) \mord \heap(\addrp).\flowfld$ of rule \ruleref{FVPull} by assuming only that the value $\heap(\addrp).\flowfld$ must be at least as large as the value propagated along the edge $(\addr,\addr')$ in $\heap$. Here, $\addr$ is the predecessor node from which $\addr'$ was read, as in rule \ruleref{FVPull}. We make this approximation to avoid the computation of $\heap.\outflowfld(\addrp)$ at every pull site. We omit the corresponding \code{assume} statement from \cref{fig:impl:pull}.

\paragraph{Space Invariants}
A central aspect of our view abstraction is the space invariant from which the view heap is materialized.
We encode it as an uninterpreted predicate $\aninv_T$, one for each occurring pointer type $T$.
The signature of $\aninv_T$ is $\aninv_T : \mcode{int} \times \mcode{int}^k \times \mcode{int}^2 \to \mcode{int}$.
The first argument is the address of a memory location.
The next $k$ arguments are the field values of $\addr$, assuming type $T$ has $k$ fields.
The last two arguments are the dedicated $\mfree$ and $\flowfld$ ghost fields (cf. \Cref{sec:flow-abstraction}).
All fields are of type \code{int} because pointers are mapped to \code{int}, as discussed before, and Booleans are mapped to \code{int} as well, as is standard in \texttt{C}.
Finally, the invariant evaluates to an integer indicating whether or not the given arguments are part of the space invariant.

\begin{figure}
    \begin{lstlisting}[mathescape=true,belowskip=0pt]
$\heap$(x).$\flowfld$, $\heap$(y).$\flowfld$ = $\syncd$(x, $\heap_0$(x), $\heap$(x), y, $\heap_0$(y), $\heap$(y));  $\label[line]{code:ipush:sync}$
assert($\aninv$(x, $\heap$(x).field$_1$, /* ... */, $\heap$(x).field$_k$, $\heap$(x).$\mfree$, $\heap$(x).$\flowfld$));  $\label[line]{code:ipush:invx}$
assert($\aninv$(y, $\heap$(y).field$_1$, /* ... */, $\heap$(y).field$_k$, $\heap$(y).$\mfree$, $\heap$(y).$\flowfld$));  $\label[line]{code:ipush:invy}$
$\heap_0$(x) = $\heap$(x); $\heap_0$(y) = $\heap$(y);  $\label[line]{code:ipush:save}$
\end{lstlisting}
    \caption{An idealized instrumentation template for pushing a footprint consisting of pointers \code{x} and \code{y}. The flow at \code{x} and \code{y} is recomputed and their view heap entries are pushed back into the space invariant. The template is idealized in that the view heap is not resolved down to our actual encoding (cf. \cref{sec:impl:view-heaps}).\label{fig:impl:push}}
\end{figure}

\paragraph{Pushes}
It remains to evict entries from the view heap back into the space invariant.
We refer to the program locations where this happens as \emph{push sites}.
The placement of push sites follows our formalization from \cref{sec:view-abstraction,sec:flow-abstraction}.
Whenever the entry for an address $\addr$ is no longer referenced by the view heap $\heap$, then $\addr$ is evicted.
Again, we identify the program locations where this happens statically with standard dataflow techniques.
While this may result in unnecessary early pushes, it is sufficiently precise in practice.

The instrumentation of pushes differs from the formalization in two key aspects.
First, addresses $\addr$ are not actually evicted from $\heap$; they are kept in $\heap$ and then simply ignored or overwritten by subsequent pulls.
The instrumentation of $\heap$ from \cref{sec:impl:view-heaps} guarantees that $\heap$ does not grow unboundedly.
The second difference concerns the $\sync$ operator.
While our formalization recomputes the flow for the entire view heap, our instrumentation recomputes the flow only for the actual footprint of the update, i.e., the subset of addresses that have been changed.
Additionally, newly allocated addresses are added to the footprint to ensure the footprint's interface does not change if such an address receives flow for the first time.
The flow computation for the footprint then follows \cref{sec:sync-optimization}.
We denote this optimized implementation as $\syncd$.
It takes as arguments the addresses $\addr$ that participate in the footprint as well as their field valuations $\heap_0(\addr)$ and $\heap(\addr)$ before and after the update, respectively.
It returns the new flow values for the given addresses $\addr$, or raises an assertion error if the update is not local (cf. \cref{sec:sync-optimization}).

A template for pushing two pointers \code{x} and \code{y} is given in \cref{fig:impl:push}, a generalization to any other size of footprint is as expected.
First, the flow is recomputed using $\syncd$ (\cref{code:ipush:sync}).
Then, the fields of \code{x} and \code{y} are pushed back into the space invariant (\cref{code:ipush:invx,code:ipush:invy}).
Finally, the backup $\heap_0(\mcode{x})$ is set to the new values $\heap(\mcode{x})$ and similarly for \code{y} (\cref{code:ipush:save}).
This is an optimization to elide pulls after pushing.

\subsection{Encoding View Heaps in the Stack}
\label{sec:impl:view-heaps}

The core aspect of our approach is the concept of a view heap.
So far, we treated view heaps as mathematical maps $\heap$ that provide valuations of the fields $\fld$ of addresses $\addr$.
However, using an actual, dynamic map tarnishes our goal of eliminating all heap usage in the instrumented program and achieving efficient tool support.
To address this, we encode the view heap $\heap$ into the stack.

Intuitively, this is done by mapping pointer variables, rather than their addresses, to field values and then storing those field values alongside the mapped pointer.
To develop this idea, we first convert $\heap$ into a \emph{pointer-based} view heap $\pbheap$.
The type of $\pbheap$ is $\pbheap : \vars_p \to (\flds \to \mcode{int})$.
That is, $\pbheap$ maps pointers \code{x} to a field map $\fldval : (\flds \to \mcode{int})$.
As discussed in \cref{sec:impl}, the type of all fields in the instrumented program will be \code{int}.

It is worth noting that $\pbheap$ can be viewed as a statically-sized array, since the pointer variables $\vars_p$ of the program are a finite, statically known set.
To avoid using arrays, which may not be well-supported by back-end solvers, we transform $\pbheap$ into a collection of stack variables.
For this, we introduce a stack (program) variable for every field of every pointer entry in $\pbheap$.
In other words, we associate a program variable $\mcode{x}_\fld$ with each field $\fld$ of each pointer \code{x}. Consequently, accessing $\pbheap(\mcode{x}).\fld$ translates to using $\mcode{x}_\fld$.
Overall, the rewriting strategy from \cref{sec:impl} then replaces $\heap(\mcode{x}).\fld$ with $\mcode{x}_\fld$.

\begin{figure}
    \begin{lstlisting}[mathescape=true,belowskip=0pt]
assert(x != NULL); $\label[line]{code:rpull:null}$
x_field$_1$ = havoc; /* ... */ x_field$_k$ = havoc; x_$\mfree$ = havoc; x_$\flowfld$ = havoc; $\label[line]{code:rpull:havoc}$
assume($\aninv$(x, x_field$_1$, /* ... */, x_field$_k$, x_$\mfree$, x_$\flowfld$)); $\label[line]{code:rpull:inv}$
if (x == y) {$\label[line]{code:rpull:alias_begin}$
  assume(x_field$_1$ == y_field$_1$ && /* ... */ && x_field$_k$ == y_field$_k$ &&
         x_$\mfree$ == y_$\mfree$ && x_$\flowfld$ == y_$\flowfld$); } $\label[line]{code:rpull:alias}$
assert(x_$\mfree$ == 0); $\label[line]{code:rpull:free}$
x$_0$_field$_1$$\;$=$\;$x_field$_1$; /* ... */ x$_0$_field$_k$$\;$=$\;$x$_0$_field$_k$; x$_0$_$\mfree$$\;$=$\;$x_$\mfree$; x$_0$_$\flowfld$$\;$=$\;$x_$\flowfld$; $\label[line]{code:rpull:save}$
\end{lstlisting}
    \caption{An instrumentation template for pulling pointer \code{x}, encoding the view heap as stack variables. This refines the idealized template from \cref{fig:impl:pull}. To improve the precision of this per-pointer encoding of the per-address view heap, \crefrange{code:rpull:alias_begin}{code:rpull:alias} perform a \emph{dynamic} alias analysis that reconciles the field values of aliasing pointers. This is a dynamic analysis in the sense that the back-end solver will resolve the alias check on \cref{code:rpull:alias_begin}.\label{fig:impl:pull-real}}
\end{figure}

To illustrate the approach, \cref{fig:impl:pull-real} provides an updated template of a pull site for pointer \code{x} that encodes the view heap in the stack.
The $\pnull$-check remains unchanged (\cref{code:rpull:null}).
The view heap entries for \code{x} are non-deterministically chosen by \code{havoc}-ing the field variables associated with \code{x} (\cref{code:rpull:havoc}).
In the template, these variables take the form \code{x_$\fld$}, where $\fld$ is the name of a field accessible through \code{x}, including the special fields $\mfree$ and $\flowfld$.
Next, the space invariant is assumed for the materialized fields (\cref{code:rpull:inv}).
It receives as arguments the pointer \code{x} and the values \code{x_$\fld$} of its fields.
(Ignore \crefrange{code:rpull:alias_begin}{code:rpull:alias} for the moment.)
Subsequently, \code{x} is asserted not to be freed, illustrating the use of the field variable \code{x_$\mfree$}.
Finally, a backup of the materialized fields is created to be used in pulls (\cref{code:rpull:save}).
This copy is encoded with stack variables in exactly the same way as the view heap.

\medskip
A notable drawback of the stack encoding of view heaps is a loss of precision.
Using a dynamic map $\heap$ for addresses essentially includes an alias analysis.
For two pointers, \code{x} and \code{y}, pointing to the same address $\addr$, both $\heap(\mcode{x}).\fld$ and $\heap(\mcode{y}).\fld$ would refer to the same value.
Since $\heap$ is address-based, evaluating \code{x.$\fld$} and \code{y.$\fld$} always yields the same result, because it resolves aliases.

However, with our encoding, we introduce two distinct variables, \code{x_$\fld$} and \code{y_$\fld$}.
To enhance the precision of this approach, we incorporate a dynamic alias analysis into the instrumentation process.
This analysis is dynamic in that it is part of the instrumentation and thus ultimately performed by the back-end solver.
We add it to pull sites.
Specifically, when pulling \code{x}, for each distinct pointer \code{y}, we insert a conditional check to determine if \code{x} and \code{y} are aliases.
If they are, we assume that all their fields have identical valuations.
This mechanism is illustrated on \crefrange{code:rpull:alias_begin}{code:rpull:alias} in \cref{fig:impl:pull-real}.

A similar issue arises with update commands such as \code{x.$\fld$ = 5}.
In our pointer-based encoding, this command updates \code{x_$\fld$} to $5$ but does not alter \code{y_$\fld$}.
This discrepancy affects not just precision but also soundness.
An unreflected update on \code{y_$\fld$} could allow bugs in the original program to go undetected.
To address this, we again utilize a dynamic alias analysis.
The instrumentation for updates inserts, for every pointer \code{y} distinct from \code{x}, a conditional \lstinline!if (x == y) { y_$\fld$ = x_$\fld$; }!, which updates the field variable of the aliasing pointer accordingly.

\subsection{Encoding Memory Safety}
\label{sec:impl:mem-safety}

We detail the assertions in our instrumentation that guarantee memory safety.
As discussed in \cref{sec:motivation}, a crucial aspect for establishing memory safety is maintaining an invariant that accurately captures which objects in the heap graph are reachable.
To achieve this, we employ a \emph{path counting flow}.
This flow counts the number of distinct paths that reach a given node from a set of initial addresses.
In our instrumentation, the initial addresses are those pointed to by global variables, each receiving an initial path count of $1$.
The flow is then propagated through the pointer fields of objects.
Users have the option to annotate specific pointer fields to instruct \atool{triceratops} to exclude them from the flow propagation.
Our evaluation in \cref{sec:eval} demonstrates that this flow abstraction is sufficient even for complex data structures, such as doubly-linked lists and trees.

Checking for unsafe accesses and double frees is straightforward.
To prevent unsafe accesses, we insert assertions that ensure all dereferences target non-null pointers referencing memory that has not yet been freed (which is indicated by the auxiliary $\mfree$ field).
In fact, the necessary checks are already included in the pull sites, as detailed in \cref{sec:impl:overview}.
Since we pull all variables that are dereferenced, no additional checks are required to safeguard against unsafe accesses.
To prevent double frees, the instrumentation inserts the expected assertions before every \code{free(x)} command.

\begin{figure}
    \begin{lstlisting}[mathescape=true,belowskip=0pt]
x$\;$=$\;$havoc; x_field$_1$$\;$=$\;$havoc; /* ... */ x_field$_k$$\;$=$\;$havoc; x_$\mfree$$\;$=$\;$havoc; x_$\flowfld$$\;$=$\;$havoc; $\label[line]{code:no-leak:havoc}$
assume(x != NULL); $\label[line]{code:no-leak:non-null}$
assume($\aninv$(x, x_field$_1$, /* ... */, x_field$_k$, x_$\mfree$, x_$\flowfld$)); $\label[line]{code:no-leak:inv}$
assert(x_$\flowfld$ > 0 || x_$\mfree$ == 1); $\label[line]{code:no-leak:assert}$
\end{lstlisting}
    \caption{Instrumentation template for ensuring that no memory is leaked at any point during execution.\label{fig:impl:no-leaks}}
\end{figure}

The most intricate aspect of ensuring memory safety involves checking for memory leaks. We address this by distinguishing between two types of memory: owned memory and shared memory.
Owned memory refers to memory that has been allocated using a local variable and has not yet been pushed into the space invariant.
Our instrumentation uses standard dataflow analyses to identify such memory regions.
We insert assertions to ensure that these memory regions are freed before all pointers to them go out of scope or are overwritten.
This is as expected.
The remaining, non-owned memory is shared.
Because it has been pushed to the space invariant, we can rely on it for the necessary checks.
To prevent leaks of shared memory, we assert that all addresses covered by the space invariant either continue to have flow or are freed.
The universal quantification is realized by a non-deterministic choice.
The assertions are added to the finalization of the input program, meaning they are executed at the very end of the instrumented program.

A template for performing the shared memory leak check is given in \cref{fig:impl:no-leaks}.
In \cref{code:no-leak:havoc}, an address and its field values are chosen non-deterministically.
This selection is restricted to non-$\pnull$ addresses (\cref{code:no-leak:non-null}) that satisfy the space invariant (\cref{code:no-leak:inv}).
The critical assertion is on \cref{code:no-leak:assert}.
It verifies that the chosen address has not been leaked at any point during program execution.
This assertion is comprehensive, covering all moments during execution, given that the space invariant is maintained throughout the entire execution.

\subsection{The Semantics of Allocations}
\label{sec:impl:no-realloc}

Our instrumentation of allocations adheres to the program model outlined in \cref{sec:prog-model}, which specifies that freed memory is never reallocated.
As a result, our implementation of \code{malloc} is fairly simple.
It increases a counter for the last allocated address and returns its new value.
To ensure the soundness of not reallocating memory, we follow \cite{DBLP:conf/vmcai/HazizaHMW16,DBLP:journals/pacmpl/MeyerW19,DBLP:journals/pacmpl/MeyerW20} and let our instrumentation insert assertions that
\begin{inparaenum}
  \item prevent the dereferencing of dangling pointers, and
  \item prevent the comparison of dangling pointers for equality with other pointers (except $\pnull$).
\end{inparaenum}
For the former check there is nothing to do, because it is part of pull sites and all pointers are pulled before being dereferenced.
For the latter check, we insert assertions before every pointer equality comparison to ensure that the participating pointers are non-dangling.
Observe that, in the absence of reallocations, a pointer is dangling if and only if the $\mfree$ marker of the referenced address is raised.
This observation is in fact the reason that we can have $\mfree$ markers per addresses rather than per pointer, which greatly simplifies our instrumentation.
We would need per pointer $\mfree$ markers if memory was reallocated, because an allocation only validates the receiving pointer.
All previously dangling pointers to the reallocated address remain dangling.
Hence, the allocation status of the referenced address does not help when deciding whether or not a pointer is dangling.
This intricacy justifies our design decision to avoid reallocation and check additional properties instead.

\subsection{Concurrency}
\label{sec:impl:concurrency}

While our primary results focus on sequential programs, we have developed an experimental extension of \atool{triceratops} to address concurrency.
Our concurrency model treats the program as a library of functions executed arbitrarily by an arbitrary number of threads.
To verify such programs, \atool{triceratops} employs a thread-modular proof approach \cite{DBLP:journals/cacm/OwickiG76,DBLP:conf/cav/KraglQH20,DBLP:conf/ifip/Jones83}, where the program is analyzed from the perspective of a single, isolated thread.

For this sequential proof to generalize to a concurrent proof, we must account for \emph{interference} from other threads, i.e., for the effect of actions from other threads.
This is achieved by relying on the space invariant, which now acts like a \emph{yield} invariant in~\cite{DBLP:conf/cav/KraglQH20}.
After executing an atomic command in the sequential program, the view heap is updated to a new version consistent with the space invariant, effectively re-pulling all pointers.
These update points within the proof are commonly referred to as \emph{yield} points.

To improve the precision of this model, we statically compute lock sets for each program location, which capture all locks a thread is guaranteed to hold at that point.
These lock sets allow us to omit the re-pulling of pointers that are protected by the locks, providing an approximation of addresses that are \emph{not} subject to interference by other threads.
To ensure the soundness of this approach, in instrumentation, any updates to the fields of an address are guarded by assertions that verify that the address is covered by the lock set at that specific location.

We used this extended instrumentation as a proof of concept.
Surprisingly, it enabled us to verify complex structures like lock-based linked lists and trees, substantiating the merit of our approach.
Notably, it generalizes much more easily to various settings compared to traditional, hand-crafted shape analyses.



\section{Evaluation}
\label{sec:eval}


\begin{table*}%
	\newcommand{\hintmark}{\textsuperscript{\textasteriskcentered}\xspace}
	\newcommand{\rtime}[1]{$#1 \mkern+1mu s$}
	\newcommand{\rtimeYes}[1]{\rtime{#1}~\symbolYes}
	\newcommand{\rtimeNo}[1]{\rtime{#1}~\symbolNo}
	\caption{%
		Sequential benchmarks; run on Apple M1 Pro; runtime averaged over 10 runs. Split into three parts: \atool{triceratops} instrumentation, \atool{tricera} export to SMT, and \atool{eldarica} solving SMT. LoC = lines of code; LoI = lines of instrumentation. Benchmarks with \hintmark required a simple hint (creation of an alias). Symbols \symbolYes and \symbolNo indicate whether the tool produced a correct verification verdict.%
		\label{table:sequential_eval}%
	}
	\label{table:eval}%
	\center%
    \footnotesize%
	\setlength{\tabcolsep}{4pt}
	\begin{tabularx}{\textwidth}{Xccrrrrr}%
		\toprule
Benchmark & LoC & LoI & \atool{triceratops} & \atool{tricera} & \atool{eldarica} & Total & \atool{predator-hp}
\\\midrule
dll-fst-data.c & 83 & 402 & \rtime{1.41} & \rtime{1.38} & \rtime{5.08} & \rtimeYes{7.87} & \rtimeNo{0.87} \\\midrule
dll-lst-data.c & 80 & 402 & \rtime{1.44} & \rtime{1.39} & \rtime{6.81} & \rtimeYes{9.64} & \rtimeNo{0.86} \\\midrule
dll-middle-data.c & 78 & 402 & \rtime{1.64} & \rtime{1.51} & \rtime{4.70} & \rtimeYes{7.85} & \rtimeNo{0.74} \\\midrule
dll.c & 68 & 339 & \rtime{1.18} & \rtime{1.42} & \rtime{4.02} & \rtimeYes{6.62} & \rtimeYes{0.60} \\\midrule
sll-fst-data.c & 72 & 394 & \rtime{1.03} & \rtime{1.42} & \rtime{4.39} & \rtimeYes{6.84} & \rtimeYes{0.60} \\\midrule
sll-lst-data.c & 72 & 394 & \rtime{1.05} & \rtime{1.51} & \rtime{4.32} & \rtimeYes{6.88} & \rtimeYes{0.60} \\\midrule
sll-middle-data.c & 76 & 398 & \rtime{1.19} & \rtime{1.45} & \rtime{7.05} & \rtimeYes{9.69} & \rtimeYes{0.61} \\\midrule
sll-shared-after.c & 88 & 563 & \rtime{2.62} & \rtime{1.87} & \rtime{12.23} & \rtimeYes{16.72} & \rtimeYes{0.60} \\\midrule
sll-shared-before.c & 79 & 634 & \rtime{1.47} & \rtime{1.94} & \rtime{3.90} & \rtimeYes{7.31} & \rtimeYes{0.61} \\\midrule
sll-shared-sll-after.c & 117 & 650 & \rtime{6.84} & \rtime{2.35} & \rtime{6.99} & \rtimeYes{16.18} & \rtimeYes{0.61} \\\midrule
sll-shared-sll-before.c & 108 & 634 & \rtime{4.52} & \rtime{2.22} & \rtime{4.69} & \rtimeYes{11.43} & \rtimeYes{0.60} \\\midrule
sll.c & 64 & 338 & \rtime{0.89} & \rtime{1.35} & \rtime{3.69} & \rtimeYes{5.93} & \rtimeYes{0.61} \\\midrule
dll-fst-shared.c & 97 & 816 & \rtime{4.60} & \rtime{2.24} & \rtime{7.86} & \rtimeYes{14.7} & \rtimeYes{0.61} \\\midrule
dll-lst-shared.c & 98 & 816 & \rtime{4.49} & \rtime{2.13} & \rtime{7.52} & \rtimeYes{14.14} & \rtimeYes{0.60} \\\midrule
dll-middle-shared.c & 109 & 1717 & \rtime{11.43} & \rtime{3.40} & \rtime{13.94} & \rtimeYes{28.77} & \rtimeYes{0.60} \\\midrule
sll-fst-shared.c & 93 & 812 & \rtime{3.62} & \rtime{2.18} & \rtime{7.69} & \rtimeYes{13.49} & \rtimeYes{0.60} \\\midrule
sll-lst-shared.c & 92 & 812 & \rtime{3.63} & \rtime{2.05} & \rtime{7.26} & \rtimeYes{12.94} & \rtimeYes{0.60} \\\midrule
sll-middle-shared.c & 106 & 1713 & \rtime{8.95} & \rtime{3.17} & \rtime{18.12} & \rtimeYes{30.24} & \rtimeYes{0.60} \\\midrule
sll-shared-sll.c & 115 & 910 & \rtime{8.40} & \rtime{2.42} & \rtime{12.14} & \rtimeYes{22.96} & \rtimeYes{0.60} \\\midrule
dll-ins.c & 35 & 277 & \rtime{0.69} & \rtime{1.42} & \rtime{2.24} & \rtimeYes{4.35} & \rtimeYes{0.60} \\\midrule
dll-ins-del.c & 53 & 337 & \rtime{1.06} & \rtime{1.85} & \rtime{3.00} & \rtimeYes{5.91} & \rtimeYes{0.60} \\\midrule
filter-01.c & 37 & 248 & \rtime{0.46} & \rtime{1.21} & \rtime{11.27} & \rtimeYes{12.94} & \rtimeYes{0.60} \\\midrule
filter-02.c & 38 & 248 & \rtime{0.51} & \rtime{1.31} & \rtime{14.17} & \rtimeYes{15.99} & \rtimeYes{0.60} \\\midrule
list-append.c & 26 & 207 & \rtime{0.32} & \rtime{1.02} & \rtime{2.59} & \rtimeYes{3.93} & \rtimeYes{0.60} \\\midrule
list-prepend.c & 21 & 200 & \rtime{0.19} & \rtime{1.05} & \rtime{1.75} & \rtimeYes{2.99} & \rtimeYes{0.60} \\\midrule
list-ins.c & 31 & 225 & \rtime{0.41} & \rtime{1.16} & \rtime{6.21} & \rtimeYes{7.78} & \rtimeYes{0.60} \\\midrule
list-ins-del.c & 47 & 275 & \rtime{0.58} & \rtime{1.34} & \rtime{11.04} & \rtimeYes{12.96} & \rtimeYes{0.60} \\\midrule
tree-ins.c & 31 & 237 & \rtime{1.19} & \rtime{1.44} & \rtime{12.69} & \rtimeYes{15.32} & \symbolNo \\\midrule
tree-ins-del.c & 63 & 347 & \rtime{4.61} & \rtime{1.96} & \rtime{262.79} & \rtimeYes{269.36} & \symbolNo \\\midrule
dll-ins-del-unsafe.c & 55 & 329 & \rtime{1.07} & \rtime{1.99} & \rtime{3.03} & \rtimeYes{6.09} & \rtimeYes{0.60} \\\midrule
filter-02-unsafe.c & 39 & 238 & \rtime{0.48} & \rtime{1.15} & \rtime{7.89} & \rtimeYes{9.52} & \rtimeYes{0.60} \\\midrule
list-ins-del-unsafe.c & 49 & 271 & \rtime{0.58} & \rtime{1.24} & \rtime{3.85} & \rtimeYes{5.67} & \rtimeYes{0.60} \\\midrule
tree-ins-unsafe.c & 34 & 230 & \rtime{1.24} & \rtime{1.23} & \rtime{2.52} & \rtimeYes{4.99} & \rtimeYes{0.60} \\\bottomrule
	\end{tabularx}
\end{table*}

We evaluated \atool{triceratops} on a set of benchmarks that covers a variety of different shapes, including singly-linked, doubly-linked, and nested lists, as well as trees, to substantiate that our approach is capable of verifying programs where this task inherently relies on shape information.

Our benchmark set is divided into three parts.
First, we selected $18$ of the $31$ sequential examples from the \texttt{memsafety-broom}\footnote{Available at: \url{https://gitlab.com/sosy-lab/benchmarking/sv-benchmarks/-/tree/svcomp24-final/c/memsafety-broom}} subset of the SV-COMP test suite \cite{DBLP:conf/tacas/Beyer24}.
The examples chosen from \texttt{memsafety-broom} feature singly-linked lists, doubly-linked lists, and nested lists.
We excluded examples featuring circular queues, due to incompatibility with simplified locality requirement on updates (cf. \cref{sec:sync-optimization}), and those that necessitate more complex flow domains than path-counting, which is the only flow domain presently supported by \atool{triceratops}.
Second, we included 14 standard sequential shape analysis examples involving singly- and doubly-linked lists, as well as binary trees, to stress-test \atool{triceratops}. Four of these benchmarks exhibit memory safety violations. The remaining ones are memory-safe.
Lastly, our benchmark set features six memory-safe examples of concurrent singly-linked lists and binary trees.
All our benchmarks are specifically designed as test cases for memory safety verification.
As such, they non-deterministically generate dynamic structures in the heap, possible modify or traverse them, and eventually free them.

The experimental results for the sequential and concurrent examples are given in \cref{table:sequential_eval} and \cref{table:concurrent_eval}, respectively.
The tables list
\begin{inparaenum}
 	\item the name of the benchmark,
 	\item the lines of code (LoC) of the example program,
 	\item the lines of code of the instrumentation (LoI) that \atool{triceratops} produces,
 	\item the time it takes \atool{triceratops} to produce the instrumentation,
 	\item the time it takes \atool{tricera} to encode the heap-less instrumentation into Horn clauses,
 	\item the time it takes \atool{eldarica} to solve Horn clauses,
 	\item the total time it takes for our tool chain to check memory safety, and
 	\item a mark \symbolYes if the verification result is as expected and \symbolNo otherwise.
\end{inparaenum}
The experiments are conducted on an Apple M1 Pro and runtimes are averaged across $10$ runs.

We also compare the performance of \atool{triceratops} with the state-of-the-art shape analysis tool \atool{predator-hp}~\cite{DBLP:conf/tacas/PeringerSV20,DBLP:conf/hvc/HolikKPSTV16}, which is the 2024 SV-COMP gold medalist in the memory safety category \cite{DBLP:conf/tacas/Beyer24}.
The verification for \atool{predator-hp} is reported in the last column of \cref{table:sequential_eval} together with an indication whether the tool produced the expected output.

Our benchmarks show that our new approach is capable of verifying intricate memory safety tasks.
As expected, the runtimes are slower than those of a fine-tuned shape analysis tool like \atool{predator-hp}.
However, the easy adaptability of our approach allows us to verify, for instance, sequential and concurrent tree benchmarks that are beyond \atool{predator-hp}'s capability, such as tree benchmarks. We believe that an extension of \atool{predator-hp} to trees or concurrency would be highly non-trivial.
This substantiates the usefulness of arithmetizing shape analysis to leverage a broader range of verification tools.

It is worth noting that the runtimes for sequential and concurrent benchmarks are comparable. We believe this is due to the fact that materializing the space invariant results in similarly precise shape information in both the sequential and concurrent case and that the re-pulling aspect required for the concurrent instrumentation (cf. \cref{sec:impl:concurrency}) can be dealt with rather easily by the back-ends.

Finally, our analysis is over-approximate, so that we may suffer from counter-examples that are not genuine. However, from counter-examples, it can often be reconstructed \emph{why} the failing assertion was added by our instrumentation. This can reveal actual bugs in programs. While this process is not automated, we successfully used it during the development of \atool{triceratops} and during the experiments.



\section{Related Work}
\label{sec:related-work}

\subsubsection*{Shape Analysis}
Numerous other studies are related to our work, mainly in the domain of shape analysis, some of which we discuss below. For a more in-depth overview we refer the reader to~\cite{DBLP:journals/ftpl/ChangDMRR20}.

A parametric framework for shape analysis based on three-valued logic was first proposed in~\cite{DBLP:journals/toplas/SagivRW02}. This approach is generic and can theoretically be applied to any program; the effectiveness depends heavily on the chosen predicates, which must be manually tailored to the analyzed programs. The three-valued logic approach is reformulated in terms of predicate abstraction in \cite{DBLP:conf/sas/PodelskiW05}.

There is a wide range of abstract domains specialized for shape analysis.
The use of monotonic abstraction for shape analysis is proposed in~\cite{DBLP:conf/dagstuhl/AbdullaBCHJR08}. \cite{DBLP:conf/cav/BouajjaniDERS10} introduces an abstract domain for singly-linked list, augmented with domains to reason about the data in the list. \atool{memcad}~\cite{DBLP:conf/vmcai/ToubhansCR13,DBLP:journals/fmsd/IllousLR21,DBLP:conf/sas/GietRR23} is a framework that utilizes a combination of sequence and shape abstractions to analyze data structures such as lists and trees.
\cite{DBLP:conf/popl/ChangR08} presents the integration of parametric abstract domains to enable detailed heap abstractions. The approach allows for the customization of abstract domains with user-defined predicates about data properties.


\begin{table*}%
	\newcommand{\hintmark}{\textsuperscript{\textasteriskcentered}\xspace}
	\newcommand{\rtime}[1]{$#1 \mkern+1mu s$}
	\newcommand{\rtimeYes}[1]{\rtime{#1}~\symbolYes}
	\caption{%
		Concurrent benchmarks; run on Apple M1 Pro; runtime averaged over 10 runs. Split into three parts: \atool{triceratops} instrumentation, \atool{tricera} export to SMT, and \atool{eldarica} solving SMT. LoC = lines of code; LoI = lines of instrumentation. Symbols \symbolYes and \symbolNo indicate whether the tool produced a correct verification verdict.%
		\label{table:concurrent_eval}%
	}
	\center%
    \footnotesize%
	\setlength{\tabcolsep}{4pt}
	\begin{tabularx}{\textwidth}{Xcrrrrr}%
		\toprule
Benchmark & LoC & LoI & \atool{triceratops} & \atool{tricera} & \atool{eldarica} & Total
\\\midrule
list-01-conc.c &       41 &      267& \rtime{0.73} & \rtime{1.27} & \rtime{3.70} & \rtimeYes{5.70} \\\midrule
list-04-conc.c &       60 &      327 & \rtime{1.11} & \rtime{1.52} & \rtime{19.62} & \rtimeYes{22.25} \\\midrule
tree-00-conc.c &       38 &      248 & \rtime{1.01} & \rtime{1.44} & \rtime{7.71} & \rtimeYes{10.16} \\\midrule
tree-01-conc.c &       49 &      287 & \rtime{1.88} & \rtime{1.68} & \rtime{27.29} & \rtimeYes{30.85} \\\midrule
tree-02-conc.c &       35 &      264 & \rtime{1.38} & \rtime{1.61} & \rtime{16.44} & \rtimeYes{19.43} \\\midrule
tree-03-conc.c & 72 & 417 & \rtime{4.51} & \rtime{2.45} & \rtime{293.56} & \rtimeYes{300.52} \\\midrule
	\end{tabularx}
\end{table*}

An abstract domain for lists is used in the tool \atool{predator}~\cite{DBLP:conf/sas/DudkaPV13}, a verifier for memory safety of low-level programs. 
It uses symbolic memory graphs (SMG) to encode sets of heap configurations, and can handle
various forms of lists. \atool{predator} is sound, but has no mechanism to
detect spurious counterexamples (false alarms). \atool{predator-hp}~\cite{DBLP:conf/tacas/MullerPV15} addresses this by running additional instances of \atool{predator} in parallel with no heap abstraction, allowing the tool to detect most false alarms. The main weakness of \atool{predator} is the lack of support for
non-pointer data. It also cannot handle trees and skip-lists in a sound way,
and does not support concurrency. The approach of SMG is also used by \atool{cpachecker}~\cite{DBLP:conf/tacas/BaierBCJKLLRSWW24}, a model checker for C programs, in combination with symbolic execution \cite{DBLP:conf/kbse/0001018} for the purpose of checking unbounded memory safety.

Shape analysis using forest automata is explored in~\cite{DBLP:conf/cav/HolikLRSV13} and~\cite{DBLP:journals/acta/AbdullaHJLTV16}, which employ abstractions capable of handling data structures as complex as skip-lists.

There are several type-based approaches to shape analysis.
The paper \cite{DBLP:conf/esop/KuruG19} details a type system designed for safe memory reclamation in concurrent programming, emphasizing the use of Read-Copy-Update (RCU). RCU enhances performance by enabling lock-free read access, while the type system ensures that memory is managed correctly across concurrent operations.
In~\cite{DBLP:conf/vmcai/NicoleLR22}, the authors discuss the integration of linear types with SMT-based verification to enhance memory reasoning. The work in~\cite{DBLP:journals/pacmpl/LiLZCHPH22} proposes using types to summarize the heap's byte-level layout.

Several studies employ the more expensive global reasoning, but deliberately stay within decidable logics in order to still enable automation~\cite{DBLP:conf/cav/ItzhakyBINS13, DBLP:conf/popl/LahiriQ08, DBLP:conf/popl/MadhusudanQS12, DBLP:conf/cade/WiesMK11}.

\subsubsection*{Separation Logic and the Flow Framework}
Separation logic~\cite{DBLP:conf/csl/OHearnRY01, DBLP:conf/lics/Reynolds02} is a widely used formalism for reasoning about both memory safety and functional properties.
Early versions of \atool{predator}~\cite{DBLP:conf/cav/DudkaPV11} were based on separation logic with inductive predicates.
To enable reasoning about dynamic data structures in the heap, the key idea in separation logic is to support compositional reasoning. 
By separating the heap into smaller regions, properties for different heap regions can be expressed locally, such that modifications to one region do not invalidate properties about other regions.

Finding intuitive local proofs in separation logic can be challenging.
A prevalent obstacle is relating global properties to local ones. 
Several important properties of data structures depend on non-local information and are not directly compatible with modular reasoning.
For example, the property that a graph of nodes in a heap has the shape of a tree cannot be merely described using local invariants on individual nodes.

The flow framework~\cite{DBLP:journals/pacmpl/KrishnaSW18,DBLP:conf/esop/KrishnaSW20,DBLP:conf/tacas/MeyerWW23}, which we build on in this paper, endows the (heap) graph with a quantity called flow that allows global properties to be specified in terms of node-local invariants by referring to that flow.
Flows are additional ghost information that are augmented to the nodes in the heap.
They are computed inductively over the graph structure using data-flow equations.
The flow framework has been used in the verification of sophisticated algorithms that are difficult to handle by other techniques, such as the Priority Inheritance Protocol, object-oriented design patterns, and complex concurrent data structures~\cite{DBLP:conf/pldi/KrishnaPSW20}.

In~\cite{DBLP:conf/esop/KrishnaSW20} the authors develop a general proof technique for reasoning about global graph properties using the flow framework.
In their approach, the user needs to define data structure invariants as node-local conditions on the flows.
The framework automatically checks that each modified heap region preserves the given invariants.  
The technique is implemented on top of the verification tool Viper.
Although checking the invariants is automatic in their tool, it still requires the user to provide the local invariants on nodes.
Similarly, \cite{DBLP:journals/pacmpl/MeyerWW22,DBLP:journals/pacmpl/0001W023,DBLP:conf/cav/MeyerOWW23} devise (semi-)automatic proof techniques based on the flow framework. While their techniques and tools can handle intricate fine-grained concurrent data structures, the user has to specify the data structure invariant from which a proof is constructed \cite{DBLP:journals/pacmpl/MeyerWW22,DBLP:journals/pacmpl/0001W023} or a full proof outline which is checked \cite{DBLP:conf/cav/MeyerOWW23}.

\subsubsection*{Constrained Horn Clauses}
Constrained Horn Clauses (CHC) represent a subset of First Order Logic (FOL) that is versatile enough to describe a wide range of verification, inference, and synthesis problems.
There exist several mature CHC-solvers, including \atool{spacer}~\cite{DBLP:conf/cav/KomuravelliGC14}, \atool{eldarica}~\cite{DBLP:conf/fmcad/HojjatR18} and \atool{golem}~\cite{DBLP:conf/cav/BlichaBS23}, with varying degrees of support for background theories.

A theory of heaps for CHCs was introduced in~\cite{DBLP:conf/smt/EsenR22}, and is used in the tool \atool{tricera}~\cite{DBLP:conf/fmcad/EsenR22}, which we apply as back-end in our implementation \atool{triceratops}. However, the instrumented programs output by \atool{triceratops} are heap-free and the theory of heaps was not used in this work.
Program heap can also be represented using the McCarthy theory of arrays~\cite{DBLP:conf/ifip/McCarthy62}, an approach used in \atool{seahorn}~\cite{DBLP:conf/cav/GurfinkelKKN15}, an automated analysis framework for LLVM-based languages. It applies data structure analysis~\cite{DBLP:conf/pldi/LattnerA05} in order to split heap into disjoint regions, then uses the theory of arrays for encoding them.

Solving CHC with arrays often necessitates the inference of quantified solutions, which can be hard for solvers. In~\cite{DBLP:conf/sas/BjornerMR13}, it is shown that quantifiers can sometimes be handled by rewriting the Horn clauses. In the same spirit, the authors in~\cite{DBLP:conf/sas/MonniauxG16} apply an abstraction to eliminate arrays to obtain array-free CHC. The space invariants approach taken by
\atool{jayhorn}~\cite{DBLP:conf/lpar/KahsaiKRS17}, which our view abstraction is inspired by, uses a similar rewriting to infer universally quantified invariants over heap objects. The \atool{jayhorn} approach is over-approximate and can yield false alarms; however, it tries to reduce their amount by utilizing several
refinements, such as adding flow sensitivity by passing additional arguments to the invariants.

\section{Conclusions}
\label{sec:conclusions}

We have presented a new automatic shape analysis method based on two reasoning principles: \emph{flow abstraction,} which reduces global properties of the heap graph to local flow equations that are required to hold for every object on the heap, and \emph{view abstraction,} for representing an unbounded number of heap objects using symbolic invariants. As our approach is implemented through a source-to-source transformation, it can be used in conjunction with different verification back-ends, and is able to leverage all data types and language features supported by the back-end tool. Our experiments show that the analysis approach covers a wide range of shapes and can even be extended to concurrent programs.

Several avenues for future work exist. At the moment, concurrency support in \atool{triceratops} is only experimental, more research  is needed to work out the details of how to analyze concurrent programs operating on linked data structures using our approach. We also plan to investigate the use of other flow domains, beyond path counting, to obtain more precise shape analysis. Lastly, the details of how to combine shape analysis with data analysis (e.g., sortedness of lists or well-formedness of search trees) remain to be investigated.

\begin{acks}
This work is funded in parts by the \grantsponsor{GS100000001}{National Science Foundation}{http://dx.doi.org/10.13039/100000001} under grant~\grantnum{GS100000001}{2304758}, by DARPA under Agreement No.~HR00112020022, by the Swedish Foundation for Strategic
    Research (SSF) under the project WebSec (RIT17-0011), by
the Swedish Research Council (VR) under the grant~2021-06327, and by the Wallenberg project UPDATE.
The first author is supported by a Junior Fellowship from the Simons Foundation (855328, SW).
\end{acks}

\bibliography{biblio,dblp}

\end{document}